# A computationally frugal open-source foundation model for thoracic disease detection in lung cancer screening programs

Niccolò McConnell[1,2,3] *, Pardeep Vasudev[1,2,3], Daisuke Yamada[1,2], Daryl Cheng[1,4], Mehran Azimbagirad[1,2], John McCabe[1,6], Shahab Aslani[1,4], Ahmed H. Shahin[1,2], Yukun Zhou[1,5], The SUMMIT Consortium†, Andre Altmann[1,5], Yipeng Hu[5], Paul Taylor[3], Sam M. Janes[4,6], Daniel C. Alexander[1,2] & Joseph Jacob[1,4] *.

1. Hawkes Institute, University College London, UK.
2. Department of Computer Science, University College London, UK.
3. Institute of Health Informatics, University College London, UK.
4. Department of Respiratory Medicine, University College London, UK.
5. Department of Medical Physics and Biomedical Engineering, University College London, UK
6. Lungs for Living Research Centre, UCL Respiratory, University College London, London, UK

†. Summit Consortium authors and affiliations listed at end of file.
*Corresponding authors. Contact email: niccolo.mcconnell.17@ucl.ac.uk, j.jacob@ucl.ac.uk.

**Low-dose computed tomography (LDCT) imaging employed in lung cancer screening (LCS) programs is increasing in uptake worldwide. LCS programs herald a generational opportunity to simultaneously detect cancer and non-cancer-related early-stage lung disease. Yet these efforts are hampered by a shortage of radiologists to interpret scans at scale. Here, we present TANGERINE, a computationally frugal, open-source vision foundation model for volumetric LDCT analysis. Designed for broad accessibility and rapid adaptation, TANGERINE can be fine-tuned off the shelf for a wide range of disease-specific tasks with limited computational resources and training data. Relative to models trained from scratch, TANGERINE demonstrates fast convergence during fine-tuning, thereby requiring significantly fewer GPU hours, and displays strong label efficiency, achieving comparable or superior performance with a fraction of fine-tuning data.**
**Pretrained using self-supervised learning on over 98,000 thoracic LDCTs, including the UK's largest LCS initiative to date and 27 public datasets, TANGERINE achieves strong performance across 14 disease classification tasks, including lung cancer and multiple respiratory diseases, while generalising robustly across diverse clinical centres. By extending a masked autoencoder framework to 3D imaging, TANGERINE offers a scalable solution for LDCT analysis, departing from recent closed, resource-intensive models by combining architectural simplicity, public availability, and modest computational requirements. Its accessible, open-source lightweight design lays the foundation for rapid integration into next-generation medical imaging tools that could transform LCS initiatives, allowing them to pivot from a singular focus on lung cancer detection to comprehensive respiratory disease management in high-risk populations.**

National lung cancer screening (LCS) programs herald a generational opportunity to identify early pre-symptomatic disease phenotypes for some of the most common chronic respiratory diseases in the world. Of participants invited to attend LCS programs only 2% will be diagnosed with lung cancer, while the remaining subjects represent a cohort enriched for the presence of diverse chronic lung diseases including chronic obstructive pulmonary disease (COPD), interstitial lung disease (ILD) and bronchiectasis [1–3]. Currently, for COPD and ILD, disease is diagnosed when symptomatic patients have perturbed lung physiology, confirmed by abnormalities in lung function tests – by which time lung damage is already established and/or advanced. In contrast, LCS programmes afford the opportunity to detect preclinical stages of airways or interstitial lung damage, where imaging abnormalities are radiologically visible despite lung function tests remaining normal. This early detection capability could allow respiratory medicine to shift from late-stage treatment to early intervention [4,5].

The potential impact on population health of improved early thoracic disease detection using LCS imaging is appreciable when considering its demonstrated cost-effectiveness in UK-based trials which have reported incremental cost-effectiveness ratios of between £8,466 and £10,069 per quality-adjusted life year gained[6,7]. Additionally, the UK National Health Service LCS program is aiming for 100% coverage of high-risk individuals (current or former smokers aged 55–74) by 2030, which will deliver nearly one million scans annually[8]. However, a critical barrier to realising the potential of LCS imaging in disease detection lies in the global shortage of radiologists. In England alone, the radiologist workforce shortfall is predicted to be 40% by 2028 [9]. Without sufficient radiologist capacity, there exists a compelling need to develop robust and scalable alternative image interpretation methods.

While traditional supervised deep learning methods have shown promise in assisting radiologists with LDCT interpretation, they typically rely on large, expertly annotated datasets [10–15]. These models often generalise poorly across clinical centres, reflecting their dependence on narrow, structured radiology datasets. Moreover, methods have often relied on patch-based approaches that risk losing contextual information and require prior knowledge of disease location for model development. Existing deep learning approaches in lung cancer screening have largely targeted focal lesions such as lung nodules, emphysema and consolidation, typically drawing on structured radiology reports [16] – these lesions taken in isolation overlook the presence of subtle parenchymal changes (such as interstitial lung abnormalities (ILAs) and mild bronchiectasis) reflecting the early stages of chronic lung diseases which require specialised radiology expertise to recognise and characterise.

Foundation models can learn from unlabelled data via self-supervised learning (SSL) [17–25], which enables the creation of pretrained models with generalisable feature extraction abilities that can then be fine-tuned for adaptation to downstream tasks like disease classification using fewer labelled data (thereby displaying label-efficiency) [26]. However, recently released CT vision foundation models suffer from important limitations. The Gemini based model [16,27] lacks publicly available pretrained weights and relies on cloud-based infrastructure to generate embeddings, which poses challenges for data privacy and limits its applicability to access-restricted or sensitive datasets. While some recent models such as M3FM [16,27] report strong performance, they utilise large-scale compute infrastructure for model pretraining and fine-tuning and the limited availability of pretrained weights can pose challenges for reproducibility and downstream benchmarking. These limitations constrain the utility of such models in research and clinical environments, where computational resources are often limited. Hence, there remains a pressing need for foundation models that are not only accurate and generalisable, but also lightweight, open-access, and computationally efficient – enabling fine-tuning with limited data and resources. Such models must support local deployment for transparency and scalability, while maintaining strong performance across a diverse range of tasks and clinical settings.

We present TANGERINE (Thoracic Autoencoder Network Generating Embeddings for Radiological Interpretation of Numerous End-tasks), a lightweight, open-source vision foundation model for volumetric thoracic LDCT analysis. Designed with frugality in mind, TANGERINE enables local, off-the-shelf, low-resource fine-tuning without reliance on cloud infrastructure – supporting widespread adoption and reproducible experimentation across the research community. The model was pretrained on over 98,000 LDCT scans from public and private datasets, including the SUMMIT[28] study and 27 public datasets spanning eight countries. TANGERINE leverages a computationally efficient 3D masked autoencoder framework [29] and a streamlined Vision Transformer (ViT) [30] architecture for volumetric representation. It enables full-scan assessment for the simultaneous detection of coexisting respiratory diseases and achieves strong performance across 14 classification tasks, including lung cancer, interstitial lung abnormalities (ILA), and multi-label disease classification. We focus on classification tasks as a scalable, data-efficient benchmark to evaluate representation quality, generalisation, and label efficiency across diverse phenotypes and domains which are core goals in foundation model development. TANGERINE demonstrates rapid convergence, robust generalisation, and high label efficiency – matching or surpassing other pretrained and non-pretrained models while requiring substantially less labelled data. These results highlight the potential for accessible foundation models to support earlier detection and more effective management of thoracic disease worldwide.

The construction and evaluation framework for TANGERINE is summarised in Fig. 1. In stage one, the model was pretrained on a curated CT dataset, including SUMMIT one of the world's largest lung cancer screening studies, and quality-assessed public datasets reviewed by two board-certified radiologists. In stage two, the model is fine-tuned to specific downstream disease tasks using labelled data and then evaluated across three settings. Pretrain-seen-distribution tasks: Fine-tuning and evaluation on dataset distributions encountered during pretraining, ensuring subject exclusion; this tests whether SSL pretraining on unlabelled data improves performance on labelled subsets. Pretrain-unseen-distribution tasks: Fine-tuning and evaluation on datasets not seen during pretraining, assessing adaptability to new domains. Domain generalisation: Fine-tuning on one dataset and evaluating on unseen datasets, simulating deployment without labelled adaptation data. We also assessed label efficiency by fine-tuning the model on limited labelled data, and evaluated the impact of pretraining dataset size and heterogeneity for downstream task performance. Details of datasets and experimental settings are provided in section Methods. TANGERINE was compared against multiple publicly available, fine-tuneable pretrained and non-pretrained models, including CT-Foundation-Gemini[27], CT-Clip[31], ViT[30], ResNet[32], Med3D-ResNet[33], and CT-Cancer-Foundation[34]. All models were fine-tuned and evaluated on the same datasets, following their respective preprocessing pipelines to ensure fair comparison. We aim to demonstrate that a lightweight, open-access model like TANGERINE can achieve competitive performance across diverse classification tasks while remaining computationally efficient and accessible for use in academic and clinical settings.

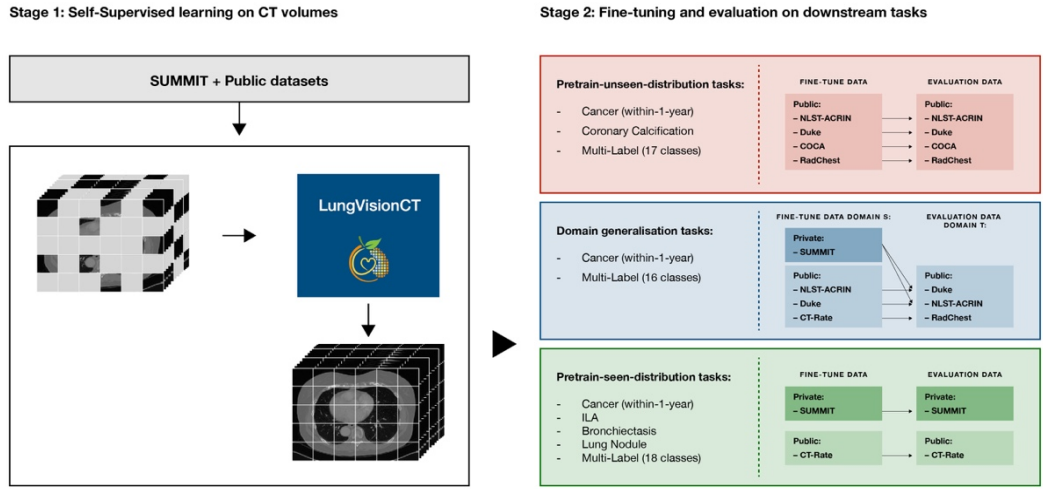

**Fig. 1 | Development and evaluation of TANGERINE.** The model undergoes two stages: **(1) Self-supervised pretraining** on LDCT scan volumes from the SUMMIT study and 27 public datasets; **(2) Supervised fine-tuning** and evaluation across three tasks: (i) Pretrain-seen-distribution, fine-tuning and testing on datasets with distributions seen during pretraining; (ii) Pretrain-unseen-distribution, fine-tuning and testing on datasets not seen during pretraining; and (iii) Domain generalisation, fine-tuning on one dataset and evaluating on a distinct target dataset unseen during pretraining or fine-tuning.

**Lung Disease Diagnosis:**

We first evaluated models on pretrain-seen-distribution tasks (Fig. 2a), where fine-tuning and evaluation data shared distributions with TANGERINE's pretraining data. Across all datasets,

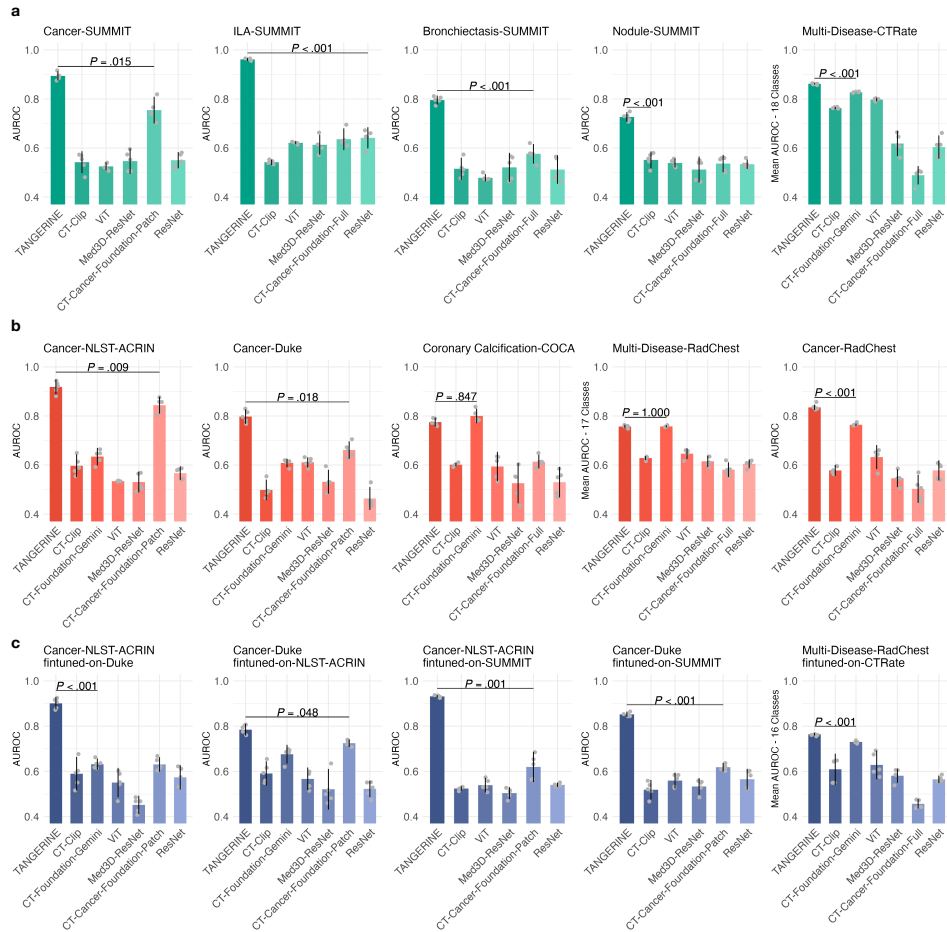

**Fig. 2 | Performance on lung disease classification**. **(a) Pretrain-seen-distribution**: Models fine-tuned and tested on datasets seen during pretraining TANGERINE. **(b) Pretrain-unseen-distribution**: Models fine-tuned and tested on datasets not seen during pretraining. **(c) Domain generalisation**: Models fine-tuned on one dataset and evaluated on a distinct target dataset unseen during pretraining or finetuning. Each model was trained with five random seeds; error bars show 95% confidence intervals, and bar centres indicate mean AUROC. Pairwise *P*-values were computed using two-tailed t-tests with Bonferroni correction (most competitive *P*-values shown, full values in Supplementary Tables 3-6). Multi-class classification results for specific disease categories are detailed in Figs. S1-3. TANGERINE consistently outperforms or matches comparison models.

TANGERINE achieved the highest AUROCs, significantly outperforming both full-scan-input and patch-based comparison models. On the within-one-year cancer classification task (SUMMIT), TANGERINE reached an AUROC of 0.894 (95% CI: 0.875–0.913), surpassing most competitive comparison model (*P*=0.015), and showed superior performance on ILA-SUMMIT, Bronchiectasis-SUMMIT, and Nodule-SUMMIT (all *P*<0.001). On Multi-Disease-CTRate, it led with the highest mean AUROC across 18 classes, significantly outperforming baselines in 15 classes (Extended Data Fig.1). We next assessed performance on pretrain-unseen-distribution tasks (Fig. 2b), where fine-tuning and evaluation dataset distributions were not encountered during model pretraining. On within-one-year cancer classification for NLST-ACRIN and Duke evaluation sets, which simulate screening cohorts with 2–3% test set cancer prevalence, TANGERINE demonstrated superior performance, significantly outperforming CT-Foundation-Gemini and CT-Cancer-Foundation-Patch. On the multi-label RadChest dataset, it matched CT-Foundation-Gemini mean AUROC across 17 classes (*P*=1.000) yet achieved significantly better performance for lung cancer (*P*<0.001) and other LCS-related conditions, including pulmonary fibrotic sequela (*P*=0.020), pleural effusion (*P*<0.001), and consolidation (*P*=0.026) (Extended Data Fig. 2). Domain generalisation was evaluated via cross-dataset evaluation where the target evaluation dataset was unseen during fine-tuning or pretraining (Fig. 2c). TANGERINE demonstrated robust generalisation capabilities, significantly outperforming CT-Foundation-Gemini and CT-Cancer-Foundation-Patch across cancer tasks. It achieved the highest mean AUROC across 16 classes in RadChest, including cardiomegaly (*P*=0.001) and pericardial effusion (*P*<0.001); however, CT-Foundation-Gemini excelled in calcification (*P*=0.001) (Extended Data Fig. 3). AUPRC results mirrored AUROC trends across evaluation settings (Extended Data Fig. 4) with TANGERINE consistently achieving the highest or statistically comparable AUPRC relative to comparison models. Grad-CAM visualisations confirmed that TANGERINE focused on pathological cancer regions during cancer prediction (Extended Data Fig. 5), and predictive entropy analyses (n=50 predictions/case) indicated consistently lower uncertainty for TANGERINE relative to non-pretrained ViT model (Extended Data Fig. 6a–c).

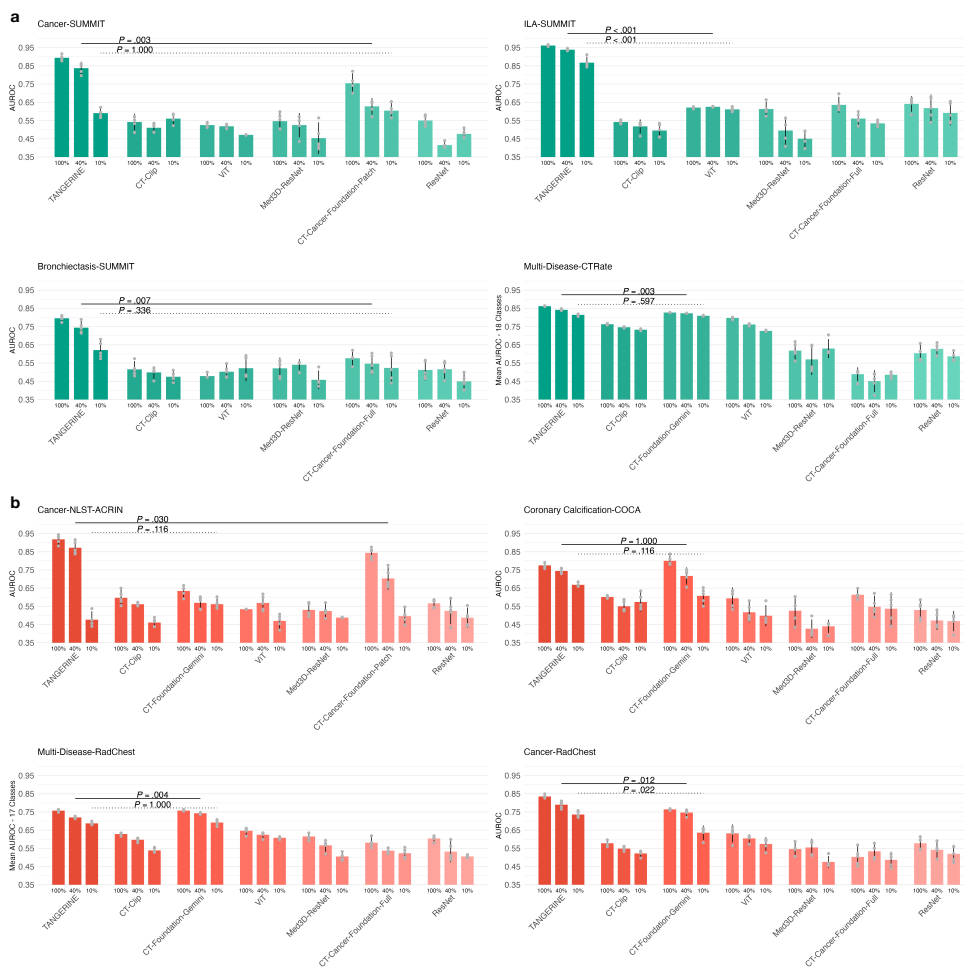

**Fig. 3 | Performance on lung disease classification and label efficiency. (a) Pretrain-seen-distribution**: Models fine-tuned and tested on datasets encountered during TANGERINE pretraining, using varying fractions of fine-tuning data (100%, 40%, 10%, noted below bars). **(b) Pretrain-unseen-distribution**: Models fine-tuned and tested on datasets not seen during pretraining, with fine-tuning data fractions indicated. Each model was trained with five random seeds; error bars show 95% confidence intervals, and bar centres indicate mean AUROC. Pairwise comparisons at 40% and 10% fine-tuning levels were assessed using two-tailed t-tests with Bonferroni correction (most competitive *P*-values shown; solid lines at 40%, dotted lines at 10%). Detailed comparisons in Supplementary Tables 8-19. Multi-class classification results are detailed in Figs. S7-8. TANGERINE displays strong label efficiency relative to comparison models.

**Training Dynamics**

As illustrated in Extended Data Fig. 7a–d, TANGERINE demonstrates rapid convergence during fine-tuning (the process of adapting a pretrained model to a specific disease task using labelled data). During this process, training loss reflects how well the model fits the training data and is used to update its weights, while validation loss estimates generalisation to unseen data. To prevent prediction based on overfit checkpoints, the model for inference is selected based on the lowest observed validation loss (early stopping). Thanks to its strong pretrained representations, TANGERINE achieves optimal validation performance within just 5–10 epochs. In contrast, while comparison models trained from scratch or with weaker pretrained initialisations eventually reduce training loss, their validation loss often remains flat or increases, indicating that these models do learn however the features they acquire do not generalise beyond the training distribution. This likely reflects suboptimal initialisation: models without strong priors must learn both low-level and high-level features from scratch, often settling into narrow, task-specific minima that fail to transfer. TANGERINE's efficiency yields practical benefits: on ILA-SUMMIT (Extended Data Fig. 7a), it reaches peak validation performance within 5–10 epochs. This corresponds to ~0.25–0.5 kWh energy consumption and 0.1–0.2 kg $CO_2$ emissions on a single NVIDIA A6000 GPU. In contrast, comparison models often require 3–5× more epochs to converge, increasing energy use and emissions substantially. These results underscore the dual value of TANGERINE pretraining: it enhances generalisation while substantially reducing training cost and carbon impact.

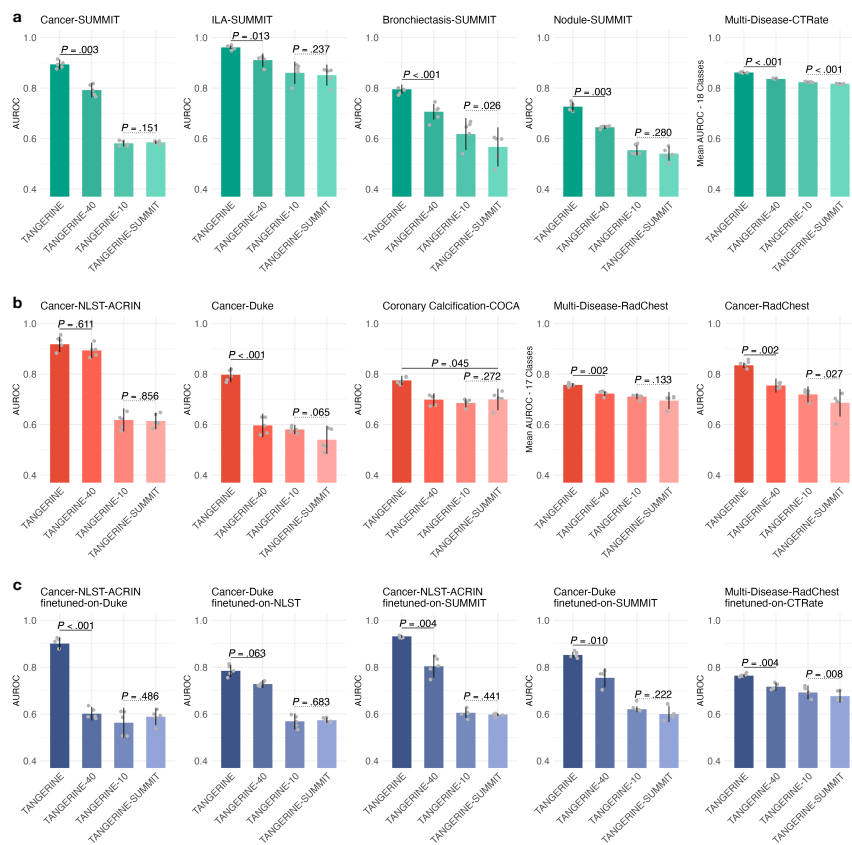

**Fig. 4 | Performance on lung disease classification with investigation of pretraining dataset size. (a) Pretrain-seen-distribution**: Models fine-tuned and tested on datasets encountered during TANGERINE pretraining. **(b) Pretrain-unseen-distribution**: Models fine-tuned and tested on datasets not seen during pretraining. **(c) Domain generalisation**: Models fine-tuned on one dataset and evaluated on datasets unseen during fine-tuning or pretraining. Pretraining dataset size was compared across TANGERINE-40 (40% pretrain data), TANGERINE-10 (10% pretrain data), and TANGERINE-SUMMIT (SUMMIT only pretrain data - homogeneous pretraining). Pairwise comparisons were made between TANGERINE (100%) and other models, as well as between homogeneous and heterogeneous pretraining (most competitive $P$-values shown; solid line comparing pretraining size; dotted lines comparing pretraining heterogeneity - full values in Supplementary Tables 20-23). Error bars denote 95% confidence intervals, and bar centres indicate mean AUROC. Multi-class classification results are in Figs. S9-11. TANGERINE performance tends to decrease with decreased pretraining dataset sizes.

**Label Efficiency:**

We assessed label efficiency by comparing model performance using 100%, 40%, and 10% of downstream fine-tuning data (Fig. 3a). On Summit-Cancer, TANGERINE fine-tuned with 40% of the data outperformed CT-Cancer-Foundation-Patch at 40% ($P=0.003$) and had no significant difference at 100% ($P=0.130$). Notably, TANGERINE with only 10% of the data on Summit-ILA exceeded all comparison models trained on 100%. On CTRate, it reached the highest AUROC at 40% relative to comparison models at all fine-tuning levels, and at 10% remained comparable to CT-Foundation-Gemini at 100%. On pretrain-unseen tasks (Fig. 3b), TANGERINE exceeded CT-Cancer-Foundation-Patch at 40% on Cancer-NLST-ACRIN ($P=0.030$), had no significant difference with CT-Foundation-Gemini at 40% and 10% on COCA, and no significant difference with CT-Foundation-Gemini on Rad-Chest at 10% (underperforming at 40%). For the Rad-Chest cancer class, TANGERINE consistently outperformed all comparison models at 40% and 10% levels.

**Effect of Pretraining Size:**

TANGERINE pretrained on the entire dataset consistently achieved the highest performance across all downstream tasks, with performance declining as pretraining dataset size decreased (Fig. 4a–c). Models with reduced pretraining data also showed reduced domain generalisation (Fig. 4c). We next compared a heterogeneous, stratified sample (TANGERINE-10) to a homogeneous, single-source dataset of similar size (TANGERINE-SUMMIT). The heterogeneous model outperformed on SUMMIT-Bronchiectasis and CTRate. Although both models performed similarly on most unseen tasks, the heterogeneous model outperformed on RadChest-Cancer ($P=0.008$) and achieved higher average performance across RadChest classes (Fig. 4b). These findings underscore that scaling pretraining data improves the generalisability of learned representations across diverse downstream imaging tasks. While additional scaling may further benefit rare or complex phenotypes, the current pretraining dataset size already yields strong and consistent performance, thereby reinforcing the value of our proposed full-data pretraining pipeline.

**Frozen embeddings demonstrate strong representational power**

Use of frozen TANGERINE embeddings is illustrated in Extended Data Fig. 8. Without any end-to-end fine-tuning, TANGERINE-Frozen achieves competitive performance across a range of tasks relative to comparison models. In the pretrain-seen-distribution setting, it consistently ranks as the second most competitive approach after full fine-tuning of TANGERINE, with the exception of the Cancer task. In the pretrain-unseen-distribution setting, TANGERINE-Frozen ranks third on the Coronary Calcification-COCA and Multi-Disease-RadChest tasks, and second on the Cancer-RadChest task. In the domain generalisation setting, it achieves the second-highest AUROC on the RadChest task when fine-tuned on CTRate. These results highlight the representational strength of the pretrained model's embeddings, particularly given the frugal nature of shallow MLP training. However, across all tasks, and especially for cancer-related endpoints, end-to-end fine-tuning offers performance gains. Additionally, to qualitatively assess the representational capacity learned during pretraining, Extended Data Fig. 9 shows reconstructed CT slices from heavily masked inputs. TANGERINE successfully reconstructs coherent lung anatomy across axial, coronal, and sagittal planes, capturing structures such as the bronchial tree, parenchymal texture, and pleural boundaries despite the sparsity of visible input.

**Discussion:**

We introduced TANGERINE, a computationally frugal, open-source foundation model for thoracic imaging that demonstrates strong performance across 14 disease classification tasks. The model exhibited high label efficiency and robust generalisation, often surpassing comparison pretrained and non-pretrained models across a range of evaluation settings, including within-distribution, out-of-distribution, and domain-generalisation tasks.

Large-scale lung cancer screening initiatives offer an unprecedented opportunity to detect early, often pre-symptomatic manifestations of chronic respiratory conditions [4,5]. LDCT captures imaging abnormalities that precede physiological decline, offering a critical window for early intervention, although interpretation at scale is hindered by global radiologist shortages. Additionally, most imaging criteria for lung disease are grounded in late-stage manifestations. Radiology is only now beginning to define imaging features that reflect early pathophysiological changes in diseases such as COPD and pulmonary fibrosis. Clinical and pharmaceutical efforts to intervene earlier are hampered by the lack of power in existing trials[35]. TANGERINE's ability to detect early imaging phenotypes, such as interstitial lung abnormalities (ILAs), could help could support identification and enrichment of patient cohorts for trials targeting preclinical disease. As diagnostic definitions shift with emerging evidence, CT tools must be repeatedly updated – a process well suited to label-efficient, and swiftly fine-tuneable models[1,2]. However, realising these clinical opportunities requires models that are not only accurate but also adaptable, data-efficient, and accessible to researchers and clinicians alike, even in resource-constrained settings.

A core strength of TANGERINE lies in its computational frugality which is achieved during both pretraining and fine-tuning. The model employs a masked autoencoding strategy in which only 25% of the input volume is processed by the encoder, reducing memory and compute demands relative to alternative self-supervised approaches [36]. This enabled pretraining on just four NVIDIA A6000 GPUs, in contrast to recent foundation models that rely on hundreds of GPUs and industrial-scale infrastructure (refer to Methods for pretraining computational requirements) [27]. During fine-tuning, TANGERINE demonstrates rapid convergence, completing training in just a few epochs and requiring significantly less GPU compute relative to comparison models. Its powerful feature extraction yields strong label efficiency, achieving performance on par with or superior to investigated models using only a fraction of annotated data.

Recent work has shown that the escalating compute requirements for training large foundation models have effectively priced out most academic labs, raising concerns about the erosion of reproducibility and innovation outside elite industry settings [37,38]. By designing TANGERINE with frugality at its core–and enabling rapid fine-tuning on a single GPU–it directly addresses this critical bottleneck. Its fully open-source framework supports local deployment and reproducible experimentation, lowering the barrier to entry for the broader research community. Beyond fine-tuning, we also demonstrate that frozen TANGERINE embeddings offer a strong and lightweight alternative, yielding competitive performance across multiple tasks when used with a shallow MLP classifier. This supports use cases where training resources are limited or real-time inference on low-power devices is desired. The ability of the frozen representations to capture disease-relevant features without end-to-end tuning reflects the strength of the self-supervised pretraining. As further evidenced by the reconstruction results (Extended Data Fig. 9), TANGERINE has learned to encode semantically rich volumetric lung anatomy–including structures such as the bronchial tree and parenchyma–underscoring the representational power of the learned embeddings, which likely contribute to both its downstream performance and label efficiency during fine-tuning.

Furthermore, unlike models that rely on cropped patches or region-level supervision, TANGERINE operates directly on full, downsampled CT volumes without requiring prior localisation of disease or manual bounding boxes. This allows for label-efficient training in real-world settings where detailed annotations are unavailable, and supports holistic reasoning across the entire thoracic cavity. By analysing the full spatial context, TANGERINE can detect diffuse or coexisting abnormalities, such as early interstitial lung changes or emphysema, that may be missed by patch-based pipelines. While some approaches [27] use multi-scale patching to capture both local and global features, TANGERINE enables end-to-end inference with a streamlined architecture that avoids dependence on anatomical priors, making it well-suited to broad screening and retrospective cohort studies. Beyond classification, TANGERINE's general-purpose representations and architectural simplicity make it well-suited for transfer to other clinically important tasks, such as lesion segmentation, disease detection, structured report generation, and quantitative scoring. These applications represent natural extensions of the current work and would further validate the model's versatility as a foundation model. By releasing all code, weights, and data pipelines, we aim to support the community in adapting TANGERINE to such diverse downstream tasks, thereby laying the groundwork for scalable, reproducible innovation in high-dimensional medical imaging. Additionally, by releasing the 3D MAE framework as open source, we aim to empower researchers to adopt similar pretraining paradigms across modalities such as brain MRI, CT angiography, and whole-body PET, catalysing progress in cardiovascular imaging, systemic disease management, and beyond.

We benchmarked TANGERINE against several widely used pretrained models, including CT-Foundation-Gemini, CT-CLIP, Med3D, and CT-Cancer-Foundation, as well as non-pretrained ViT and ResNet baselines. While each comparison model has its own strengths, such as CT-Cancer-Foundation's high performance on cancer detection and CT-Foundation-Gemini's overall robustness, TANGERINE consistently delivered strong performance across diverse disease types and domains. Notably, it achieved this while remaining computationally lightweight and fully open-access. We also observed clear differences in training dynamics. Despite convergence of training loss, the poorly performing models exhibited non-decreasing validation loss curves, indicating that the features learned failed to generalise beyond the training set. This is likely due to suboptimal initialisation, where the model overfits to low-level or task-specific features. In contrast, TANGERINE's strong pretrained representations provide an effective initialisation point for fine-tuning, enabling both rapid convergence and the acquisition of robust, transferable features. These results suggest that foundation models built with efficiency, accessibility, and generalisability in mind can serve as practical, high-impact tools for real-world clinical research, especially in settings with limited annotation or computational resources.

Despite the aforementioned advantages, TANGERINE has limitations that warrant further attention. Downsampling the input scans, while reducing computational overhead, may result in the loss of fine-grained diagnostic details in certain contexts. Future work may explore hybrid architectures or selective high-resolution input strategies to mitigate this trade-off. Furthermore, the public lung CT pretraining dataset predominantly represents a limited subset of demographic groups, potentially restricting model generalisability. Expanding datasets to include more diverse populations may further promote equitable performance across different clinical contexts. Looking ahead, prospective multi-centre studies and collaborations with healthcare providers are critical to validating TANGERINE's clinical utility and ensuring compliance with regulatory standards.

Foundation models like TANGERINE are poised to drive breakthroughs in the next wave of early detection tools and preventive care. As lung cancer screening programs scale and their clinical scope expands beyond cancer alone, tools that can flexibly adapt to emerging disease phenotypes will be indispensable. TANGERINE's open-source availability, strong generalisation performance, and low barriers to entry offer a pathway toward scalable, reproducible, and more inclusive medical AI, where innovation is no longer gated by annotation or computational infrastructure constraints. As datasets expand and pretraining methods evolve, TANGERINE illustrates how self-supervised learning can lay the foundation for personalised, preventive, and globally accessible healthcare solutions.

**Acknowledgements:**

We would like to thank Tristan Clark and Jamie O'Connor for providing access to computing resources and assisting with the execution of model pretraining on the UCL Computing Cluster.

**Funding**: This research and JJ was funded in whole or in part by the Wellcome Trust [227835/Z/23/Z ] and the NIHR UCLH Biomedical Research Centre, UK. NM acknowledges funding from the UKRI Centre for Doctoral Training (CDT) in AI-enabled Healthcare Systems, supported by UKRI grant [EP/S021612/1]. The SUMMIT Study is funded by GRAIL LLC. through a research grant awarded to S.M.J. as Principal Investigator. S.M.J. is supported by CRUK programme grant (EDDCPGM\100002), and MRC Programme grant (MR/W025051/1). SMJ receives support from the CRUK Lung Cancer Centre of Excellence (C11496/ A30025) and the CRUK City of London Centre, the Rosetrees Trust, the Roy Castle Lung Cancer foundation, the Longfonds BREATH Consortia, MRC UKRMP2 Consortia, the Garfield Weston Trust and University College London Hospitals Charitable Foundation. SMJ's work is supported by a Stand Up To Cancer-LUNGevity- American Lung Association Lung Cancer Interception Dream Team Translational Research Grant and Johnson and Johnson (grant number: SU2C-AACR-DT23-17 to S.M. Dubinett and A.E. Spira). Stand Up To Cancer is a division of the Entertainment Industry Foundation. Research grants are administered by the American Association for Cancer Research, the Scientific Partner of SU2C.This work was partly undertaken at UCLH/UCL who received a proportion of funding from the Department of Health's NIHR Biomedical Research Centre's funding scheme.

**Author Contributions:**

NM, DCA, and JJ contributed to the conception and design of the work. NM and JJ developed the methodology and experimental design. NM, PV, DY, DC, JM, SA, MA, AHS, and JJ contributed to the data acquisition and organisation. NM contributed to the technical implementation. PV, DY, DC, and JJ provided the clinical inputs to the research. NM, YH, AA, DCA, and JJ contributed to the evaluation pipeline of this work. The SUMMIT Consortium, and SMJ provided the SUMMIT data. All authors contributed to the drafting and revising of the manuscript.

**Ethics Statement:**

Written consent was obtained from all participants in the SUMMIT study following determination of study eligibility. Ethical approval was obtained from an NHS Research Ethics Committee (17/LO/2004) and the NHS Health Research Authority's Confidentiality Advisory Group (18/CAG/0054).

**Competing Interests:**

JJ declares consultancy fees from Boehringer Ingelheim, F. Hoffmann-La Roche, GlaxoSmithKline, NHSX; Advisory Boards for Boehringer Ingelheim, F. Hoffmann-La Roche;Lecture fees from Boehringer Ingelheim, F. Hoffmann-La Roche, Takeda; Grant Funding: from GlaxoSmithKline, Wellcome Trust, Microsoft Research, Gilead Sciences; Patents: UK patent application numbers 2113765.8 and GB2211487.0
SMJ has received fees for advisory board membership in the last three years from Bard1 Lifescience. He has received grant income from GRAIL Inc. He is an unpaid member of a GRAIL advisory board. He has received lecture fees for academic meetings from Cheisi and Astra Zeneca. His wife works for Astra Zeneca.

**Data and Materials Availability:**

The datasets used for pretraining are publicly available, with citations provided in Supplementary Table 1**.** Access to these datasets can be obtained through the respective sources as detailed in the table. The SUMMIT dataset, however, is subject to restricted access and is not publicly available. For the downstream tasks, the publicly available datasets used in this study are listed in Supplementary Table 2**,** along with their corresponding citations and access details. Datasets requiring institutional agreements or restricted access are also noted accordingly.
The code used in this study is provided as part of the submission and model weights are available upon request during the review process. The code is made publicly available on GitHub at [https://github.com/niccolo246/mae_reconstruction]. Further details on implementation and usage are included in the repository documentation.

**Summit Consortium Authors:**

Sam M Janes[1], Jennifer L Dickson[1], Carolyn Horst[1], Sophie Tisi[1], Helen Hall[1], Priyam Verghese[1], Andrew Creamer[1], Thomas Callender[1], Ruth Prendecki[1], Amyn Bhamani[1], Chuen Khaw[1], Mamta Ruparel[1], Monica L. Mullin[1], Tanya Patrick[1], Allan Hackshaw[2], Anne-Marie Hacker[2], Esther Arthur-Darkwa[2], Samantha L Quaife[3], Arjun Nair[4], Anand Devaraj[5], Kylie Gyertson[4], Vicky Bowyer[4], Ethaar El-Emir[4], Judy Airebamen[4], Alice Cotton[4], Kaylene Phua[4], Elodie Murali[4], Simranjit Mehta[4], Janine Zylstra[4], Karen Parry-Billings[4], Columbus Ife[4], April Neville[4], Paul Robinson[4], Laura Green[4], Zahra Hanif[4], Helen Kiconco[4], Ricardo McEwen[4], Dominique Arancon[4], Nicholas Beech[4], Derya Ovayolu[4], Christine Hosein[4], Sylvia Patricia Enes[4], Jane Rowlands[4], Sheetal Karavadra[4], Aashna Samson[4], Urja Patel[4], Fahmida Hoque[4], Hina Pervez[4], Sofia Nnorom[4], Moksud Miah[4], Julian McKee[4], Mark Clark[4], Jeannie Eng[4], Fanta Bojang[4], Claire Levermore[4], Anant Patel[6], Sara Lock[7], Alan Shaw[7], Rajesh Banka[8], Angshu Bhowmik[9], Ugo Ekeowa[10], Chris Valerio[11], William M Ricketts[12], Neal Navani[4], Ali Mohammed[12], Terry O'Shaughnessy[12], Charlotte Cash[6], Magali Taylor[4], Samanjit Hare[6], Tunku Aziz[12], Stephen Ellis[12], Anthony Edey[13], Graham Robinson[14], Alberto Villanueva[15], Hasti Robbie[16], Elena Stefan[10], Charlie Sayer[17], Nick Screaton[18], Navinah Nundlall[4], Lynsey Gallagher[4], Andrew Crossingham[4], Thea Buchan[4], Tanita Limani[4], Kate Gowers[1], Kate Davies[1], John McCabe[1], Joseph Jacob[19], Mehran Azimbagirad[19], Burcu Ozaltin[19], Tania Anastasiadis[20], Andrew Perugia[21], James Rusius[21], Geoff Bellingan[4], Maureen Browne[4], Eleanor Hellier[4], Catherine Nestor[4].

**Summit Author Affiliations:**

1. Lungs for Living Research Centre, UCL Respiratory, University College London, London
2. CRUK & UCL Cancer Trials Centre, University College London, London
3. Centre for Cancer Screening, Prevention, Detection and Early Diagnosis, Wolfson Institute of Population Health, Barts & The London School of Medicine and Dentistry, Queen Mary University of London, London
4. University College London Hospitals NHS Foundation Trust, London
5. RBHT, Imperial College
6. Royal Free London NHS Foundation Trust, London
7. Whittington Health NHS Trust, London
8. Barking, Havering and Redbridge University Hospitals NHS Trust, Essex
9. Homerton University Hospital Foundation Trust, London
10. The Princess Alexandra Hospital NHS Trust, Essex
11. North Middlesex University Hospital NHS Trust, London
12. Barts Health NHS Trust, London
13. North Bristol NHS Trust, Bristol
14. Royal United Hospitals Bath NHS Foundation Trust, Bath
15. Surrey and Sussex Healthcare NHS Trust, Surrey
16. King's College Hospital NHS Foundation Trust, London
17. University Hospitals Sussex NHS Foundation Trust, Sussex
18. Royal Papworth Hospital NHS Foundation Trust, Cambridge
19. Satsuma Lab, Centre for Medical Image Computing (CMIC), London
20. Tower Hamlets Clinical Commissioning Group, London
21. Noclor Research Support, London

**References:**

1. de Koning, H. J. *et al.* Reduced Lung-Cancer Mortality with Volume CT Screening in a Randomized Trial. *New England Journal of Medicine* **382**, (2020).
2. Gatsonis, C. A. *et al.* The national lung screening trial: Overview and study design. *Radiology* **258**, (2011).
3. de Mattos, J. N. *et al.* Computed tomography on lung cancer screening is useful for adjuvant comorbidity diagnosis in developing countries. *ERJ Open Res* **8**, (2022).
4. Ruparel, M. *et al.* Evaluation of cardiovascular risk in a lung cancer screening cohort. *Thorax* **74**, (2019).

5. Balata, H. *et al.* Targeted lung cancer screening selects individuals at high risk of cardiovascular disease. *Lung Cancer* **124**, (2018).
6. Hinde, S. *et al.* The cost-effectiveness of the Manchester 'lung health checks', a community-based lung cancer low-dose CT screening pilot. *Lung Cancer* **126**, (2018).
7. Field, J. K. *et al.* UK Lung Cancer RCT Pilot Screening Trial: Baseline findings from the screening arm provide evidence for the potential implementation of lung cancer screening. *Thorax* **71**, (2016).
8. Cancer Research UK. Lung cancer screening. (2024).
9. The Royal College of Radiologists. *Clinical Radiology Workforce Census 2023*. https://www.rcr.ac.uk/news-policy/policy-reports-initiatives/clinical-radiology-census-reports/ (2023).
10. Walstra, A. N. H. *et al.* Feasibility of AI as first reader in the 4-IN-THE-LUNG-RUN lung cancer screening trial: impact on negative-misclassifications and clinical referral rate. *Eur J Cancer* **216**, 115214 (2025).
11. Lancaster, H. L. *et al.* MA02.05 Inter-Reader Agreement of AI Versus Human Readers in Baseline LDCT Lung Nodule Classification; A UKLS Trial Dataset Sub-Study. *Journal of Thoracic Oncology* **19**, S58 (2024).
12. Gayap, H. T. & Akhloufi, M. A. Deep Machine Learning for Medical Diagnosis, Application to Lung Cancer Detection: A Review. *BioMedInformatics* vol. 4 Preprint at https://doi.org/10.3390/biomedinformatics4010015 (2024).
13. Ardila, D. *et al.* End-to-end lung cancer screening with three-dimensional deep learning on low-dose chest computed tomography. *Nat Med* **25**, (2019).
14. Javed, R. *et al.* Deep learning for lungs cancer detection: a review. *Artif Intell Rev* **57**, (2024).
15. Weiss, J. *et al.* Deep learning to estimate lung disease mortality from chest radiographs. *Nat Commun* **14**, (2023).
16. Niu, C. *et al.* Medical multimodal multitask foundation model for lung cancer screening. *Nat Commun* **16**, 1523 (2025).
17. Moor, M. *et al.* Foundation models for generalist medical artificial intelligence. *Nature* **616**, (2023).
18. Wang, X. *et al.* A pathology foundation model for cancer diagnosis and prognosis prediction. *Nature* (2024) doi:10.1038/s41586-024-07894-z.
19. Kondepudi, A. *et al.* Foundation models for fast, label-free detection of glioma infiltration. *Nature* (2024) doi:10.1038/s41586-024-08169-3.
20. Xiang, J. *et al.* A vision–language foundation model for precision oncology. *Nature* (2025) doi:10.1038/s41586-024-08378-w.
21. Fu, X. *et al.* A foundation model of transcription across human cell types. *Nature* (2025) doi:10.1038/s41586-024-08391-z.
22. Lu, M. Y. *et al.* A visual-language foundation model for computational pathology. *Nat Med* **30**, 863–874 (2024).
23. Zhang, K. *et al.* A generalist vision–language foundation model for diverse biomedical tasks. *Nat Med* (2024) doi:10.1038/s41591-024-03185-2.
24. Tiu, E. *et al.* Expert-level detection of pathologies from unannotated chest X-ray images via self-supervised learning. *Nat Biomed Eng* **6**, (2022).
25. Zhou, Y. *et al.* A foundation model for generalizable disease detection from retinal images. *Nature* **622**, (2023).
26. Huang, S. C. *et al.* Self-supervised learning for medical image classification: a systematic review and implementation guidelines. *npj Digital Medicine* vol. 6 Preprint at https://doi.org/10.1038/s41746-023-00811-0 (2023).
27. Yang, L. *et al.* Advancing Multimodal Medical Capabilities of Gemini. (2024).
28. Dickson, J. L. *et al.* Uptake of invitations to a lung health check offering low-dose CT lung cancer screening among an ethnically and socioeconomically diverse population at risk of lung cancer in the UK (SUMMIT): a prospective, longitudinal cohort study. *Lancet Public Health* **8**, (2023).
29. He, K. *et al.* Masked Autoencoders Are Scalable Vision Learners. in *Proceedings of the IEEE Computer Society Conference on Computer Vision and Pattern Recognition* vols 2022-June (2022).
30. Dosovitskiy, A. *et al.* AN IMAGE IS WORTH 16X16 WORDS: TRANSFORMERS FOR IMAGE RECOGNITION AT SCALE. in *ICLR 2021 - 9th International Conference on Learning Representations* (2021).
31. Hamamci, I. E. *et al.* Developing Generalist Foundation Models from a Multimodal Dataset for 3D Computed Tomography. (2024).
32. He, K., Zhang, X., Ren, S. & Sun, J. Deep residual learning for image recognition. in *Proceedings of the IEEE Computer Society Conference on Computer Vision and Pattern Recognition* vols 2016-December (2016).
33. Chen, S., Ma, K. & Zheng, Y. Med3D: Transfer Learning for 3D Medical Image Analysis. (2019).

34. Pai, S. *et al.* Foundation model for cancer imaging biomarkers. *Nat Mach Intell* **6**, (2024).
35. Martinez, F. J. *et al.* Treatment Trials in Young Patients with COPD and Pre-COPD Patients: Time to Move Forward. *Am J Respir Crit Care Med* **205**, (2021).
36. Yue, X. *et al. Understanding Masked Autoencoders From a Local Contrastive Perspective*.
37. Bolón-Canedo, V., Morán-Fernández, L., Cancela, B. & Alonso-Betanzos, A. A review of green artificial intelligence: Towards a more sustainable future. *Neurocomputing* **599**, (2024).
38. Besiroglu, T. *et al. The Compute Divide in Machine Learning: A Threat to Academic Contribution and Scrutiny?* (2024).
39. National Lung Screening Trial Research Team. Data from the National Lung Screening Trial (NLST). *The Cancer Imaging Archive* (2013).
40. A Wang *et al.* Duke Lung Cancer Screening Dataset 2024. (2024).
41. Draelos, R. L. *et al.* Machine-learning-based multiple abnormality prediction with large-scale chest computed tomography volumes. *Med Image Anal* **67**, (2021).
42. Stanford AIMI Center. COCA: Coronary Calcium and Chest CTs. *Stanford AIMI Center*.
43. Caron, M. *et al.* Emerging Properties in Self-Supervised Vision Transformers. in *Proceedings of the IEEE International Conference on Computer Vision* (2021). doi:10.1109/ICCV48922.2021.00951.
44. Liu, Z. *et al. A 3D Multimodal Optical Coherence Tomography Foundation Model for Retinal and Systemic Diseases with Cross-Cohort and Cross-Device Validation*.
45. Kingma, D. P. & Ba, J. L. Adam: A method for stochastic optimization. in *3rd International Conference on Learning Representations, ICLR 2015 - Conference Track Proceedings* (2015).
46. Google Research. Taking medical imaging embeddings 3D. https://research.google/blog/taking-medical-imaging-embeddings-3d/ (2024).
47. Chen, T., Kornblith, S., Norouzi, M. & Hinton, G. A simple framework for contrastive learning of visual representations. in *37th International Conference on Machine Learning, ICML 2020* vols PartF168147-3 (2020).
48. Selvaraju, R. R. *et al.* Grad-CAM: Visual Explanations from Deep Networks via Gradient-Based Localization. *Int J Comput Vis* **128**, (2020).

**Methods:**

**Datasets for developing TANGERINE:**

TANGERINE was developed using a curated collection of 98,588 CT volumes (approximately 25.2 million slices after pre-processing) from both private and public sources. The private dataset SUMMIT[28] constitutes 12.3% of the pretraining dataset. Conducted between December 2018 and May 2023, SUMMIT aimed to detect lung cancer early among at-risk Londoners and support blood test development for cancer detection. The remaining 87.7% of the pretraining dataset was derived from 27 public datasets, which were quality controlled by two board-certified radiologists, the largest of which came from CTRate[31] (47.0%) and National Lung Screening Trial–Lung Screening Study (NLST-LSS)[2] (31.2%). CTRate pretraining dataset constitutes 46,330 reconstructed 3D chest CT volumes derived from 25,692 scans of 21,304 unique patients, paired with corresponding radiology reports. These non-contrast chest CT volumes were acquired between May 2015 and January 2023 at Istanbul Medipol University Mega Hospital. NLST-LSS[2], the Lung Screening Study (LSS) component of the NLST, comprises data collected from 10 centres across the United States. NLST was a landmark study conducted from August 2002 through to April 2004 aimed at evaluating the efficacy of low-dose CT scans compared to chest X-rays for lung cancer screening in high-risk populations, involving over 53,000 participants aged 55–74 with significant smoking histories. Details of all dataset sources are provided in Supplementary Table 1. To the best of our knowledge, the pretraining dataset constitutes the most geographically and demographically diverse collection of publicly available thoracic CT scans assembled to date for training a deep learning model, comprising data from 27 sources across eight countries. The combined dataset reflects a wide diversity of imaging devices, and geographical regions, including the United States, United Kingdom, Turkey, China, Russia, Iran, Norway, Netherlands and Italy. Scans from SUMMIT were collected using state-of-the-art low-dose CT systems, while public datasets encompass clinical and research imaging setups from multiple regions. Together, these datasets provide a robust and diverse foundation for developing and pretraining TANGERINE.

**Downstream task datasets:**

We evaluated the performance of TANGERINE on several disease classification datasets. Cancer-NLST- American College of Radiology Imaging Network (ACRIN)[2,39] includes baseline 2,722 chest CT scans originating from 23 centres within the ACRIN network, distinct from NLST-LSS, with cases labelled positive if lung cancer was diagnosed within-one-year of the scan date. Cancer-Duke (USA)[40] comprises 1,104 screening chest CT scans collected between 2015 and 2021 at Duke University Health System. Nodule annotations were semi-automatically generated using a deep learning nodule detection algorithm, refined based on radiology reports, or manually verified by a fellowship-trained cardiothoracic radiologist - positive cases represent 6% of this dataset. Multi-Disease RadChest (USA)[41], collected at Duke University between 2012 and 2017, contains 3,630 publicly available non-contrast chest CT volumes with various annotated classes for which we selected annotations for 17 classes which were chosen so as to align with classes in CTRate thereby enabling domain generalisation evaluations (as detailed in[31]). CTRate (Turkey)[31], is a multi-label dataset containing 49,370 total chest CT scans labelled for 18 classes. For CTRate, we used the recommended train-test split with the training set (46,330 scans) utilised for pretraining TANGERINE, while test set (3,040 scans) was kept separate to avoid data leakage (in line with[31]). COCA (USA)[42], a dataset of 211 chest CT scans from Stanford University, includes annotations for coronary calcification, with cases labelled positive if calcification was present. From the SUMMIT (UK)[28] dataset, we evaluate subsets including Cancer-Summit (3,741 scans labelled by board-certified radiologists, with positive cases defined as lung cancer diagnoses within-one-year), ILA-Summit for interstitial lung abnormalities (920 scans labelled by two board-certified radiologists), Bronchiectasis-Summit for bronchiectasis (870 scans labelled by two board-certified radiologists), and Nodule-Summit (800 scans, where nodules were identified using a CADe system and verified by board-certified radiologists). We note SUMMIT tasks are utilised for binary classification tasks as opposed to multi-label like RadChest or CTRate. Since SUMMIT's distribution was seen during pretraining, it is considered in the pretrain-seen-distribution task dataset. However, the specific subsets used for fine-tuning were drawn from a different partition unseen during pretraining. Together, these datasets represent a diverse set of challenges, from various clinical centres with varying disease class and demographic distributions.

For model performance evaluation, most datasets were split into training, validation, and test sets in a ratio of approximately 50:10:40. For the SUMMIT cancer dataset an approximately 50:10:40 for split positive cases was utilised. A 60:10:30 was utilised for NLST-ACRIN and Duke cancer datasets due to low positive training cases. To simulate real-world screening conditions, we subsampled the NLST-ACRIN cohort to include only participants with a confirmed lung cancer diagnosis within one year of their baseline LDCT, and sampled an appropriate number of negative cases to achieve a clinically realistic test set prevalence of approximately 2–3%. While this resulted in relatively few positives in the validation set, model selection was based on validation loss, and all performance metrics were evaluated on larger, independent test sets. For example, the NLST-ACRIN-Cancer dataset includes 45 positive cases in the training set, 9 in the validation set, and 18 in the test set (1.96% prevalence). The Duke-Cancer dataset contains 38 positive cases in the training set, 7 in the validation set, and 18 in the test set (prevalence 2.9%). This setup also provides a realistic benchmark of model performance under data-constrained conditions, reflecting the challenges of many real-world applications where positive cases are rare and annotation is limited. The SUMMIT-Cancer dataset includes 156 positive cases in the training set, 26 in the validation set, and 78 in the test set (prevalence 2.6%). For CTRate we split the original training set into training and validation using an

85:15 split and kept the original provided test set for comparable performance to the original work [31]. The training set was used to optimise model parameters, while the validation set was used to monitor training convergence and select the best model checkpoint. The test set was used to evaluate the final model checkpoint and assess performance. For domain generalisation, models fine-tuned on the source training set (using the aforementioned split) were tested on the entirety of the target dataset to evaluate domain generalisation. This approach ensures performance is assessed both within the source domain (seen-pretrain-distribution and unseen-pretrain-distribution tasks) and across unseen target domains (domain generalisation). Details of downstream task datasets are provided in Supplementary Table 2.

**Dataset preprocessing and augmentation for model pretraining:**

All volumes were resampled to a uniform size of 256×256×256. To preserve each scan's physical dimensions, the new voxel spacing was dynamically computed based on the original voxel spacing and image dimensions, using the formula: new_spacing = (original_size × original_spacing) / new_size. Essential spatial metadata, including the origin, orientation, and direction, were preserved during this process. Following resampling, all processed volumes were visually inspected by two board-certified radiologists to exclude corrupted files, non-thoracic anatomy, or incomplete thoracic coverage, ensuring a high-quality dataset for model training (see Supplementary Table 1). Importantly, scans with variable slice thickness or imaging artefacts were retained to preserve the diversity and realism of real-world clinical imaging data. CT voxel intensities were clipped to a Hounsfield Unit range of [-1200, 800] focusing on the clinically relevant intensity range, and normalised to a range of [0, 1] using min-max scaling. To optimise computational efficiency during pretraining, data augmentation was limited to random flipping along the sagittal and axial planes, maintaining anatomical fidelity while introducing sufficient variability into the dataset.

**TANGERINE architecture**

Our model builds upon the masked autoencoder framework, incorporating several modifications to enable the processing of 3D CT volumes. Inspired by the success of RetFound [25], we adopted a masked autoencoding (MAE) [29] strategy for pretraining. Compared to contrastive methods such as DINO[43], MAE offers improved computational efficiency by encoding only a subset of visible patches, while avoiding the need for negative sampling and heavy augmentation pipelines. This aligns with our focus on frugal, scalable model development. Unlike the original masked autoencoder, which processes 2D image patches, TANGERINE processes 16 × 16 × 16 sub-volumes as input. To process volumetric data, the input embedding layer was replaced with a 3D patch embedding module, which partitions the CT volume into 3D patches and encodes them as feature vectors. The positional embeddings were extended to 3D sine-cosine encoding, preserving spatial relationships across three dimensions and ensuring effective modelling of the volumetric data. The architecture consists of a large Vision Transformer (ViT-large) encoder with 24 Transformer blocks and an embedding vector size of 1,024, and a Transformer based decoder with 8 Transformer blocks and an embedding vector size of 512. The decoder reconstructs 3D volumes by reintroducing masked dummy sub-volumes and projecting the embeddings back into the 3D space via a linear transformation followed by sigmoid activation to produce reconstructed voxel intensities. In contrast to large-scale, cloud-based architectures such as Gemini [27], TANGERINE employs a more frugal design, and is able to be deployed entirely on local infrastructure to reduce costs and accommodate privacy requirements in medical settings. Following the classic MAE paradigm, our encoder processes only unmasked patches, drastically reducing memory usage and improving speed (alternative methods[44] apply masking at the decoder stage and thus process all patches in the encoder, incurring higher costs) – empirically, including all patches in the encoder increased memory usage by 50% and was nearly 6× slower. This efficiency is further aided by a small decoder-to-encoder ratio, which minimises computational overhead while maintaining rich representation learning. The model was trained with a masking ratio of 0.75 for the input sub-volumes, over 400 epochs with a 20-epoch warm-up phase. We adopted a 75% masking ratio based on prior 2D MAE literature [23,29] and small-scale preliminary experiments on a subset of the data. Although these early results were indicative rather than definitive, a full ablation over masking ratios was deemed computationally prohibitive. We therefore selected 75% as a practical default that balances compute efficiency and representational challenge, consistent with our focus on frugality and accessibility. The ADAM [45] optimiser was used with an effective batch size of 256 (64 samples per GPU across four GPUs) and an initial learning rate of $1\times10^{-4}$. The loss function was mean squared error (MSE), aimed at optimising the reconstruction of 3D CT volumes. An example visualisation is of our 3D masked autoencoder framework applied to thoracic imaging is shown in Extended Data Fig. 10.

**Adaptation to downstream tasks:**

For downstream tasks, only the encoder (ViT-large) of the foundation model is retained, while the decoder is discarded. The model contains approximately 312 million parameters. The encoder processes 3D CT volumes by first dividing them into a grid of 3D patches (each corresponding to a fixed spatial sub-volume), which are embedded into 1024-dimensional feature vectors. These patch embeddings, combined with the positional embeddings, serve as inputs to the Vision Transformer. The class token, a learnable parameter appended to the input sequence, captures global contextual information during training by attending to all patch tokens across the CT volume. The patch tokens, representing localised spatial details, which in our model variant were aggregated using a global average pooling operation to summarise their contributions. This aggregation ensures that both local and global information is retained in the final representation. The class token and the averaged patch tokens are then concatenated, resulting in a 2048-dimensional feature vector (1024 from the class token and 1024 from the pooled patch tokens), which forms the high-level feature representation used for downstream tasks (this differs from the vanilla ViT which normally uses only class tokens for the final prediction). These features are passed to a linear layer whose output size corresponds to the number of classification classes (e.g., 1 for Cancer- NLST-ACRIN and 17 for RadChest). Fine-tuning is performed end-to-end, with a linear-layer head used to predict class probabilities – an example visualisation is shown in Extended Data Fig. 11. For datasets with imbalanced classes, a weighted cross-entropy loss function is employed, where the positive class weight is calculated as the ratio of positive to negative samples to ensure balanced contributions during training. The model is trained for up to 200 epochs, with an artificial batch size of 12, and a learning rate initialised at $1\times10^{-4}$. The first 10 epochs use a warm-up schedule to gradually increase the learning rate, followed by a cosine annealing schedule to reduce it progressively to $1\times10^{-6}$. A layer-wise learning rate decay of 0.75 is applied during fine-tuning, with earlier layers receiving smaller gradient updates to preserve pretrained features while allowing deeper layers to adapt to the task. For the Cancer-NLST-ACRIN and Cancer-Duke datasets, we employed a classification head comprising two linear layers. The 2048-dimensional embeddings were first passed through a hidden layer with 64 nodes, LeakyReLU activation, and batch normalisation, followed by the final output layer. This deeper architecture proved necessary as direct classification with a single linear layer resulted in suboptimal performance. Clinically realistic augmentation were applied during fine-tuning, including rotation, scaling, as well as limited Gaussian noise, Gaussian smoothing, and contrast adjustments. No augmentations were applied during inference or evaluation, ensuring that test-time predictions reflect authentic CT image characteristics. Model checkpoints were selected based on the lowest validation loss and used for evaluation on test sets. To assess the quality of TANGERINE's learned representations without end-to-end fine-tuning, we extracted frozen embeddings from the pretrained encoder and trained a shallow artificial neural network (ANN) classifier using these features. The ANN architecture consisted of two fully connected layers with hidden dimensions [128, 32], followed by a sigmoid output layer for binary classification. As a lightweight approach, this strategy supports the frugality objective of our work by enabling strong downstream performance without requiring end-to-end backpropagation through the encoder.

**Comparison models:**

We compared our model to several baseline architectures, all of which were fine-tuned on the downstream tasks, with data preprocessing aligned to the strategies specified in their original pretraining protocols. ViT: The Vision Transformer architecture used in TANGERINE was fine-tuned from scratch, employing He initialisation. CT-Clip: Employs a vision transformer-based encoder, originally developed for a vision-and-language framework, which was pretrained using contrastive SSL to match lung CT volumes with corresponding radiology reports on the CTRate dataset [31]. Input voxel values were clipped between [-1000, 1000] normalised to [-1, 1] , and resized to a uniform voxel spacing of 0.75 mm ($x$, $y$) and 1.5mm ($z$), followed by centre-cropping or padding to achieve a resolution of 480 × 480 × 240. CT-Foundation-Gemini: Based on the Google Gemini architecture pretrained on 657,719 CT scans from three U.S private hospitals, this model outputs 2048-dimensional embeddings via Google's cloud-based API[27]. While public datasets were evaluated with this model, we could not assess its performance on the SUMMIT dataset due to data-sharing restrictions. The embeddings were fine-tuned using the equivalent multilayer perceptron (MLP) as used for frozen TANGERINE embeddings which was based on the original author recommended hyperparameters[46]. ResNet-50: A vanilla ResNet-50 model was trained from scratch with He initialisation[32]. Med3D-ResNet, a ResNet-50 model pretrained on medical image segmentation tasks (Med3D) was adopted, using the encoder as a feature extractor[33]. Inputs were clipped between -1200 and 800 and normalised by the mean and standard deviation of each volume. CT-Cancer-Foundation is a ResNet-50 pretrained via SimCLR on 50×50×50 tumour patches[34,47]. We implemented two variations: CT-Cancer-Foundation-Full, processes entire volumes as input, CT-Cancer-Foundation-Patch, a patch-based model for lung cancer classification, predicts cancer probabilities for overlapping patches (50% overlap) within a scan, with the highest probability representing the scan-level prediction - this model was trained on positive cancer patches containing malignant nodule as well as benign nodules and non-nodule patches from negative cases. Preprocessing for the CT-Cancer-Foundation model normalised voxel values by clipping between [−1024 , 2048] and min-max normalising. We note all models were trained for up to 200 epochs to ensure training loss convergence, with downstream evaluation based on the best validation loss checkpoint.

**Computational resources :**

We utilised four NVIDIA A6000 GPUs (48 GB each) to train TANGERINE, with the full pretraining process requiring approximately 25 days (1.5 hours per epoch over 400 epochs). This resulted in an estimated energy consumption of ~720 kWh and a corresponding carbon footprint of ~288 kg $CO_2$ (based on 4 GPUs × 300 W × 1.5 hr × 400 epochs, assuming 0.4 kg $CO_2$/kWh). This is substantially lower than the estimated footprint of recent foundation models trained on large-scale infrastructure. For instance, M3FM [16,27] required 192 NVIDIA V100 GPUs (32 GiB each) for pretraining and multi-task training, over approximately 90 hours. Assuming 300 W per GPU, this yields an estimated 5,184 kWh and 2,074 kg $CO_2$ – more than 7 times higher than TANGERINE's training footprint. For downstream fine-tuning, TANGERINE required ~0.9 seconds per mini-batch on a single A6000 GPU. Inference on a CPU is feasible, requiring approximately 13.8 seconds per 256×256×256 volume and consuming around 5.55 GB of RAM (tested on 2019 MacBook with 2.6 GHz Intel Core i7 processor), suggesting that deployment in hospital settings without a GPU is possible, albeit with longer processing times. When using frozen TANGERINE embeddings, inference is limited to a single forward pass through the encoder followed by classification using a shallow ANN, substantially reducing memory and compute requirements. Training the ANN took approximately

0.0021 seconds per mini-batch on a CPU and required approximately 300MB of RAM, making it feasible for resource constrained settings or deployment on edge devices.

**Evaluation and Statistical Analysis:**

Task performance is evaluated using the area under the receiver operating characteristic curve (AUROC), a standard classification metric that reflects the model's ability to discriminate between classes across all thresholds. As our focus is on representation quality rather than thresholded clinical deployment, AUROC serves as a robust, threshold-independent measure of performance and is widely used in medical imaging studies. For multi-label tasks such as RadChest and CTRate, AUROC is calculated separately for each disease category, and the mean AUROC is reported as the average across all categories. Additionally, we report the area under the precision-recall curve (AUPRC), which better captures model performance on minority classes by emphasising precision and recall over true negatives, making it particularly informative in the presence of class imbalance. Each model is trained five times with different random seeds to account for variability introduced by the random shuffling of training data in line with[25]. The mean and standard deviation of performance metrics across the five runs are calculated, and the standard error is derived with confidence intervals (CIs) at 95% are computed using 1.96×standard error. Performance differences between TANGERINE and comparison models are assessed using two-sided t-tests, with Bonferroni correction applied for multiple comparisons. To assess label efficiency, additional t-tests were performed to compare model performance across varying amounts of fine-tuning data (100%, 40%, and 10%) as well as between individual training fractions for TANGERINE relative to comparison models trained on full dataset. The impact of pretraining dataset size was investigated by evaluating TANGERINE's performance using models pretrained on 100%, 40%, and 10% of the available pretraining data, with Bonferroni correction applied, and tests conducted between the model pretrained on 100% of the data and other models pretrained on fractions. Finally, the effect of pretraining diversity was assessed by comparing models pretrained exclusively on the SUMMIT dataset (homogeneous pretraining) with those pretrained on 10% of the combined dataset (heterogeneous pretraining), using two-sided t-tests.

**Grad-Cam Heatmaps:**

To provide interpretability for model predictions, we employed Gradient-weighted Class Activation Mapping (Grad-CAM)[48] to generate saliency maps highlighting regions of the input scans that contributed most significantly to cancer classification. Grad-CAM works by leveraging the gradients of the model's output with respect to the activations of the final convolutional or self-attention layer. In our Vision Transformer (ViT)-based architecture, hooks were introduced to capture the deep activations and gradients within the final transformer block. Specifically, the activation maps are weighted by the gradient signals corresponding to the target class, creating a spatial heatmap that represents the regions of highest contribution to the prediction. For each test case, the original CT scan slices in axial, coronal, and sagittal views are displayed alongside the overlaid Grad-CAM heatmaps. Red regions in the heatmaps indicate areas of high contribution to the model's classification, while cooler colours reflect lower contributions. Example visualisations are presented for three datasets: (A) Duke, (B) NLST-ACRIN, and (C) SUMMIT. These heatmaps demonstrate that the model localises cancerous tumour regions, showcasing anatomical consistency and alignment with clinical expectations. The combination of Grad-CAM with our ViT-based architecture provides crucial insights into the decision-making process, enabling us to validate that the model bases its predictions on clinically relevant features rather than spurious correlations. While we considered applying Grad-CAM to other disease categories (e.g., coronary artery calcification, bronchiectasis), preliminary results were visually ambiguous due to the diffuse or subtle nature of these conditions and the absence of pixel-level ground truth. To avoid misleading interpretations, we limited saliency visualisation to nodules and plan to explore more robust attribution methods for other pathologies in future work.

**Uncertainty Quantification:**

To quantify uncertainty across datasets and models, we utilised entropy as a measure of prediction confidence. Test-time augmentation (TTA) was employed to evaluate uncertainty by applying aggressive data transformations to the input images, generating n=50 predictions for each case in the test set. These transformations included Gaussian noise, Gaussian smoothing, contrast adjustment, flipping along all spatial axes, and affine transformations for rotation and translation. For each test case, the entropy of the 50 predictions was calculated and averaged to represent the uncertainty associated with that specific case. Additionally, we computed the overall average entropy across all cases within each dataset to summarise the uncertainty distribution. While we acknowledge that certain augmentations (e.g., smoothing or contrast jitter) may obscure subtle image features, our goal was not to generate clinically calibrated uncertainty estimates. Rather, the analysis was designed to enable a fair comparison of model robustness under perturbations, using the same augmentation protocol across all models.

# Extended Data

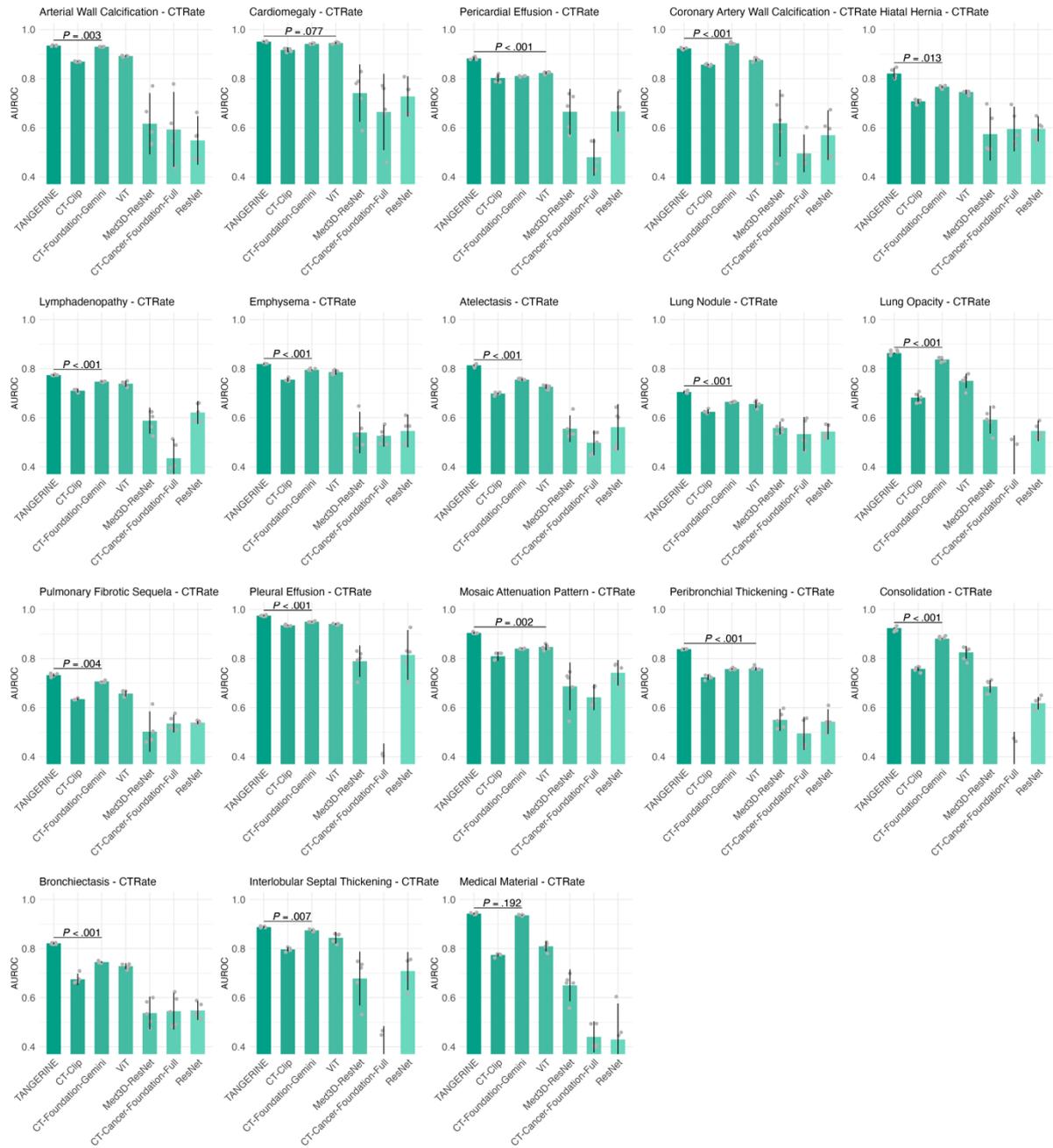

**Extended Data Fig. 1 | Performance on CTRate classes in the pretrain-seen-distribution setting.** This graph shows the performance of TANGERINE and comparison models on CTRate classes in the pretrain-seen-distribution setting. Models were fine-tuned and evaluated on CTRate, a dataset with distributions exposed to TANGERINE during pretraining. The error bars represent 95% confidence intervals (CI), and the bar centres represent the mean AUROC. Bonferroni correction was applied to adjust for multiple comparisons. The adjusted *P*-values are listed in the figure for the most competitive models (other *P*-values are provided in Supplementary Table 5). TANGERINE demonstrates high performance across CTRate classes in the pretrain-seen-distribution setting.

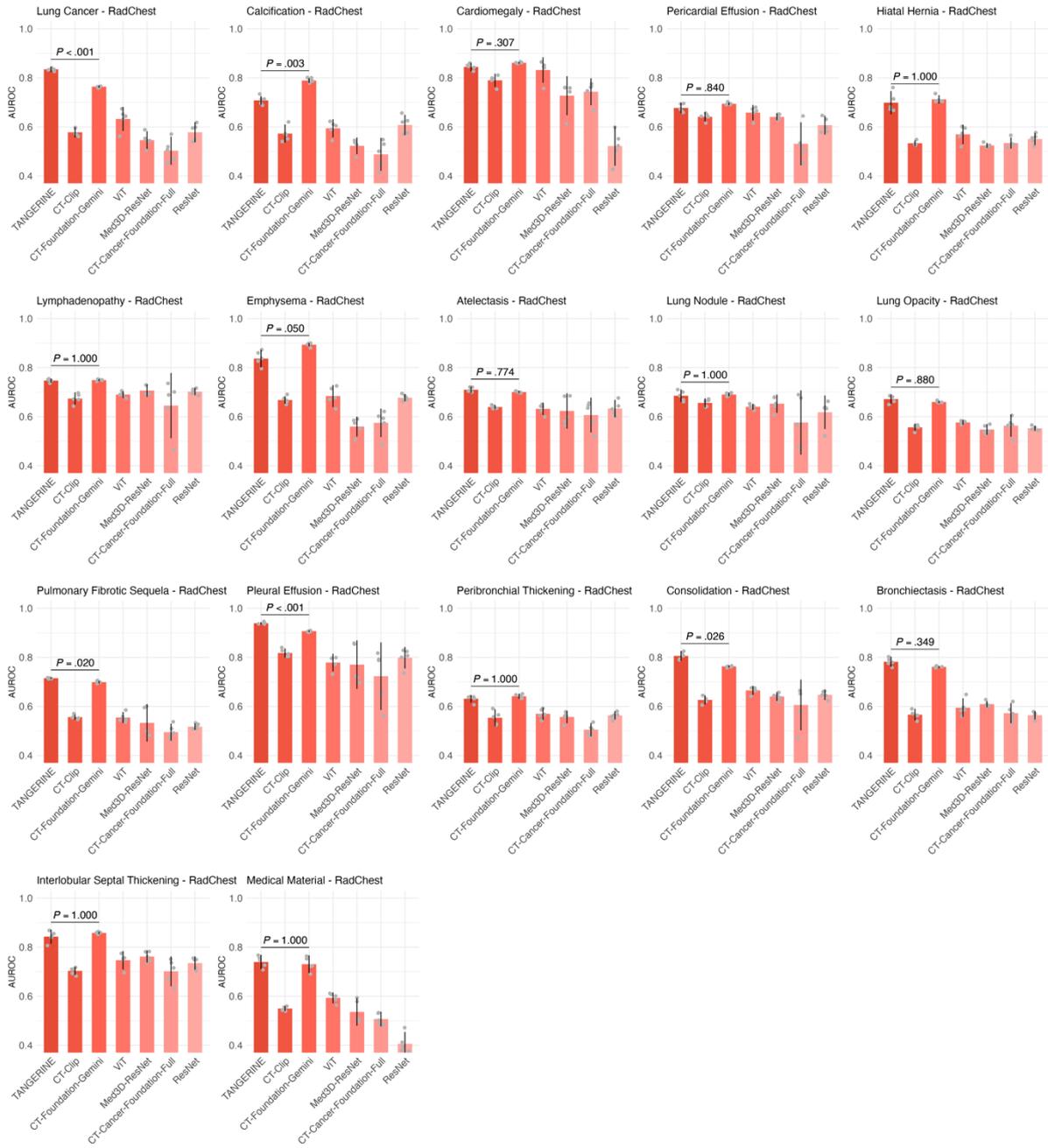

**Extended Data Fig. 2 | Performance on CTRate classes in the pretrain-unseen-distribution setting.** This graph shows the performance of TANGERINE and comparison models on CTRate classes in the pretrain-unseen-distribution setting. Models were fine-tuned and evaluated on RadChest, a dataset not seen during TANGERINE pretraining. The error bars represent 95% confidence intervals (CI), and the bar centres represent the mean AUROC. Bonferroni correction was applied to adjust for multiple comparisons. The adjusted *P*-values are listed in the figure for the most competitive models (other *P*-values are provided in Supplementary Table 5). TANGERINE demonstrates high performance across RadChest classes in the pretrain-unseen-distribution setting.

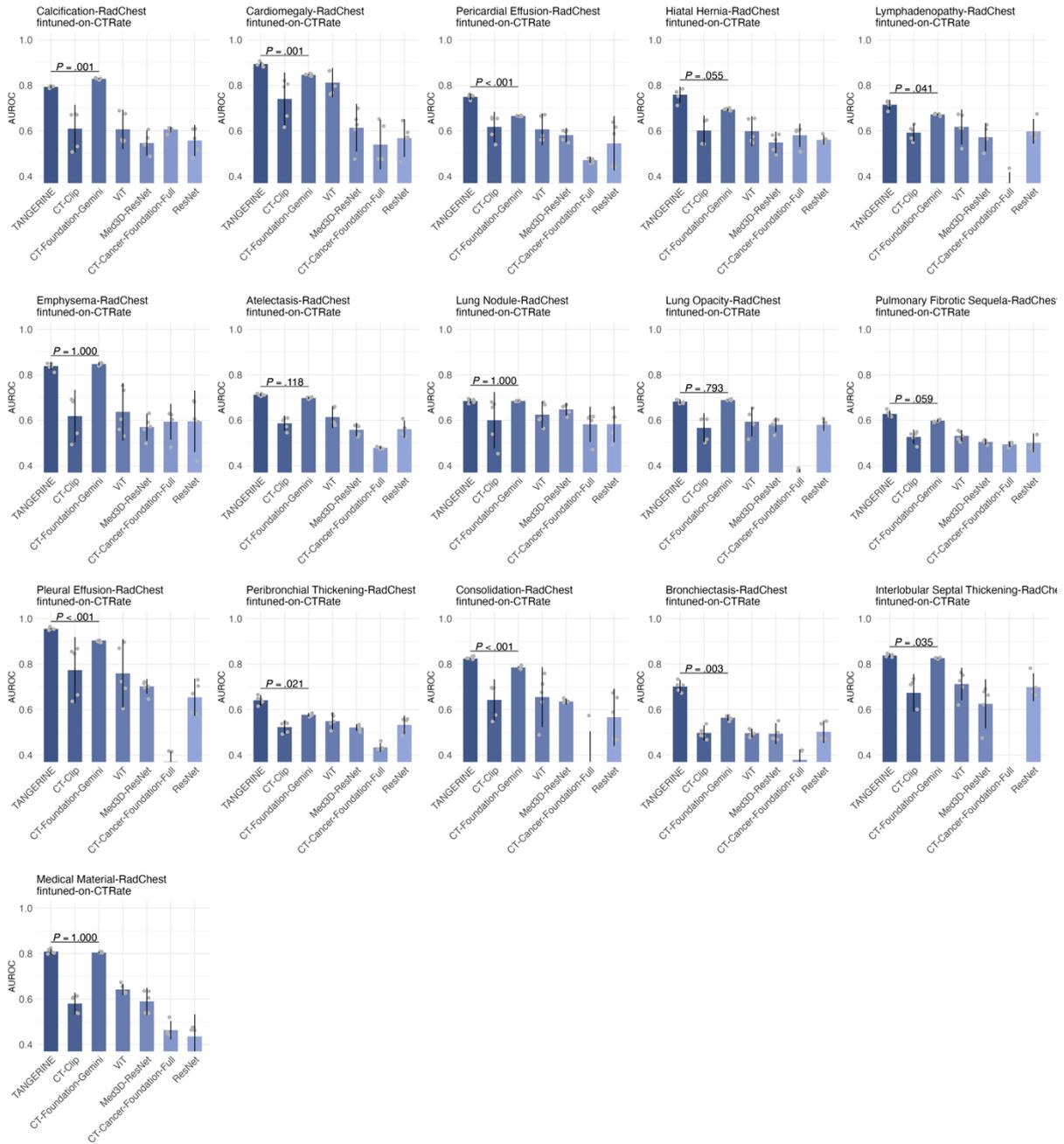

**Extended Data Fig. 3 | Performance for domain generalisation on RadChest classes fine-tuned on CTRate.** This graph shows the performance of TANGERINE and comparison models on RadChest classes corresponding to class types shared between the RadChest and CTRate datasets. Models were fine-tuned on CTRate and evaluated on RadChest to assess their ability to generalise across domains. The error bars represent 95% confidence intervals (CI), and the bar centres represent the mean AUROC. *P*-values were calculated for pairwise comparisons between TANGERINE and all other models using a two-tailed t-test, with Bonferroni correction applied to adjust for multiple comparisons. The adjusted *P*-values are listed in the figure for the most competitive models (other *P*-values are provided in Supplementary Table 8). TANGERINE demonstrates superior or comparable generalisation performance, with robust performance across the shared RadChest classes when fine-tuned on CTRate.

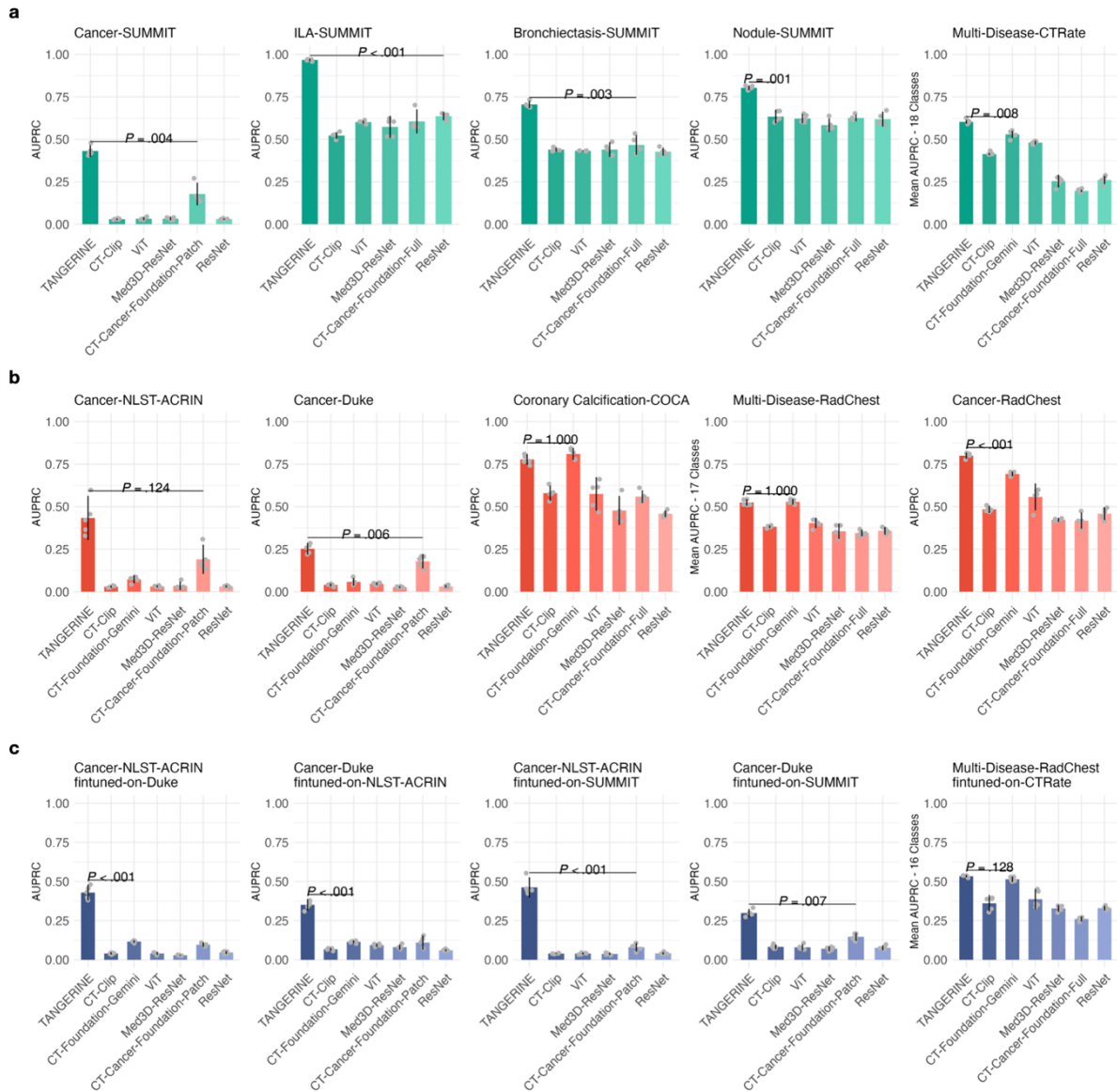

**Extended Data. 4 | Performance on lung disease classification**. **(a) Pretrain-seen-distribution**: Models fine-tuned and tested on datasets seen during pretraining TANGERINE. **(b) Pretrain-unseen-distribution**: Models fine-tuned and tested on datasets not seen during pretraining. **(c) Domain generalisation**: Models fine-tuned on one dataset and evaluated on a distinct target dataset unseen during pretraining or finetuning. Each model was trained with five random seeds; error bars show 95% confidence intervals, and bar centres indicate mean AUPRC. Pairwise *P*-values were computed using two-tailed t-tests with Bonferroni correction (most competitive *P*-values shown, full values in Supplementary Tables 7). TANGERINE consistently outperforms or matches comparison models.

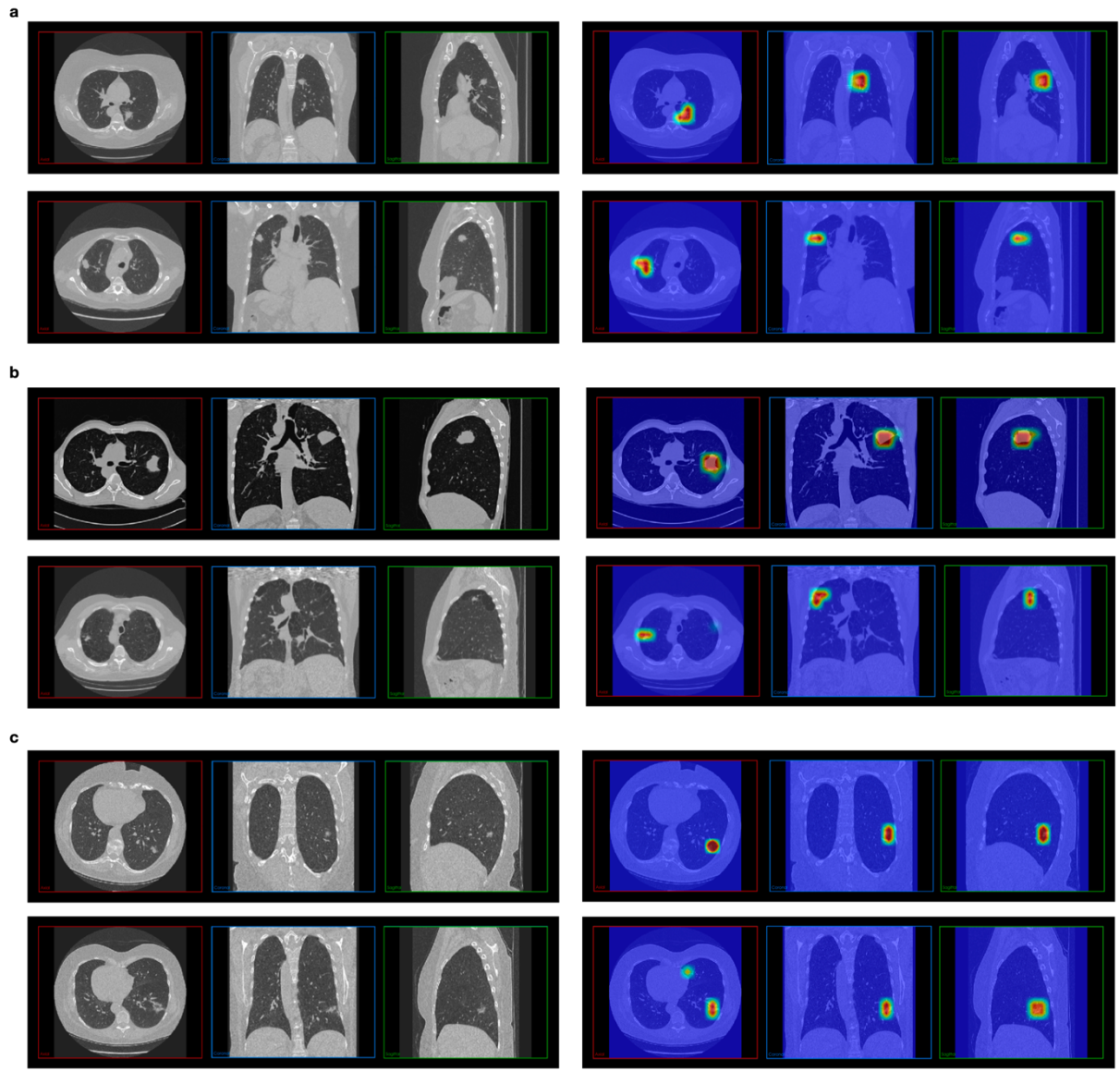

**Extended Data Fig. 5 | Cancer classification heatmaps.** Example heatmaps illustrating areas contributing to model classifications for various cancer tasks. Red indicates high contribution. On the left are the original scans in axial, coronal and sagittal order, while on the right are the overlaid Grad-CAM heatmaps for: **(a)** Duke, **(b)** NLST-ACRIN, and **(c)** SUMMIT. The highlighted cancerous tumour locations demonstrate that the model predictions are anatomically consistent.

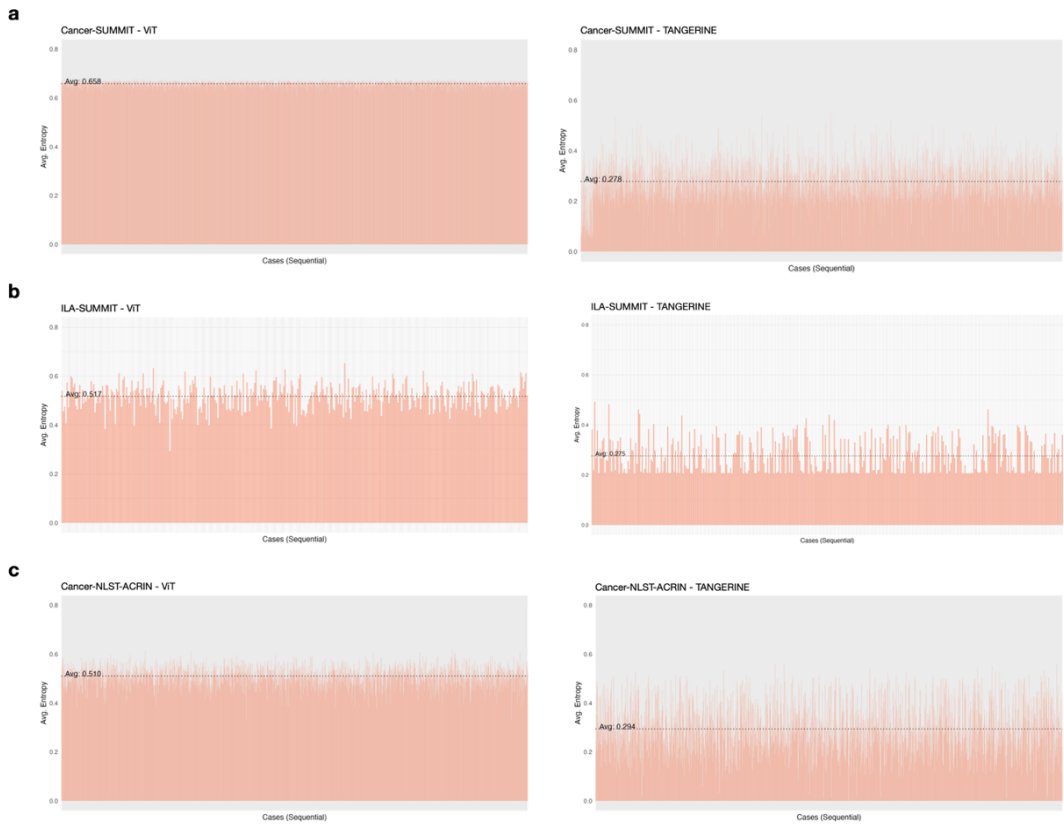

**Extended Data Fig. 6 | Uncertainty quantification across datasets and models using entropy.** For each dataset **(a) Cancer-Summit**, **(b) ILA-Summit**, and **(c) NLST-ACRIN** the bar plots show the average entropy per case after applying aggressive test-time augmentation (n=50 predictions per case). Results are presented for two models: ViT (left) and TANGERINE (right). The dotted line indicates the average entropy across all cases for each model and dataset. These visualisations highlight the distribution of uncertainty in model predictions under augmented conditions, with higher entropy reflecting greater uncertainty.

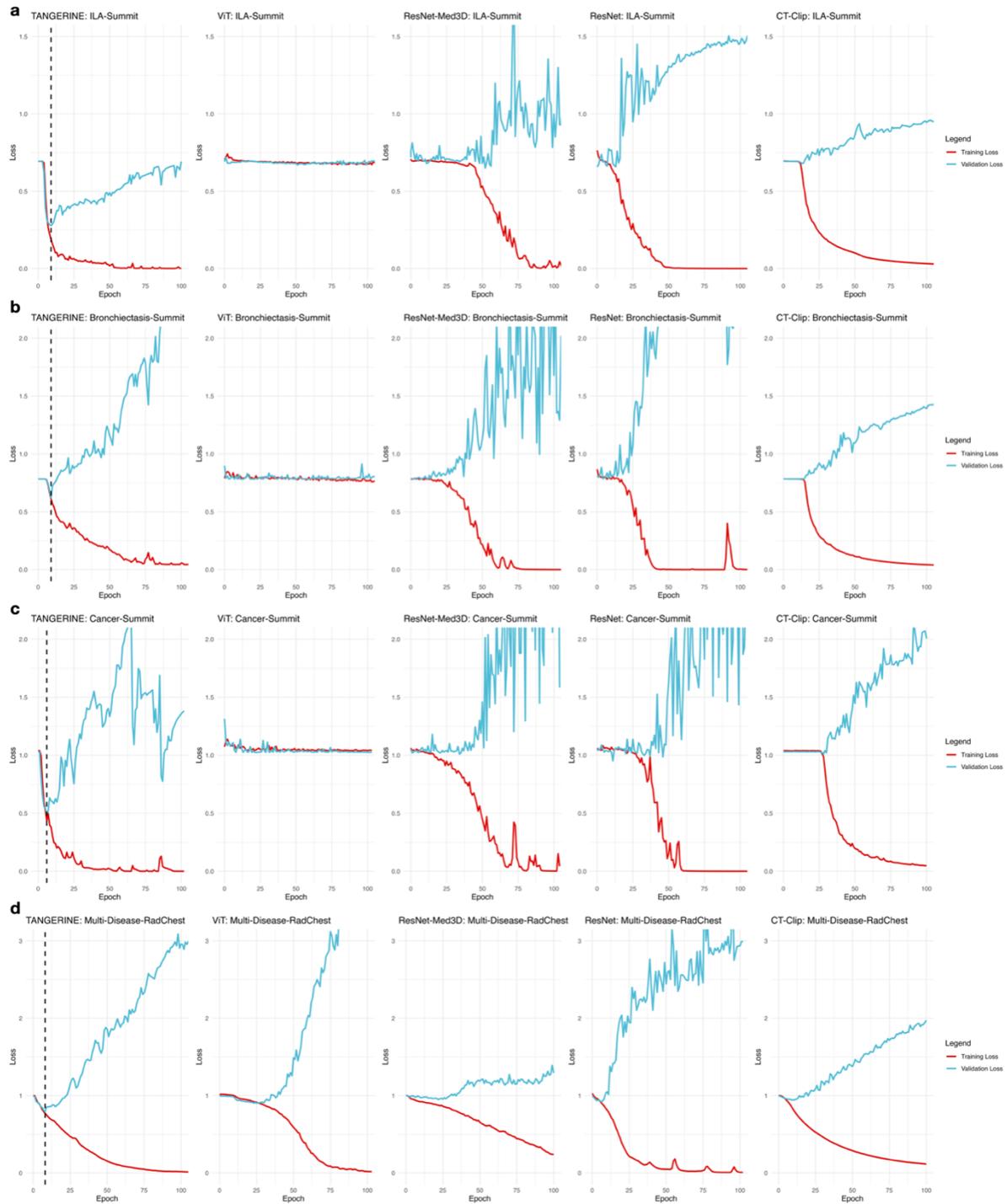

**Extended Data Fig. 7 | Training and validation loss curves for example fine-tuning tasks. (a) ILA-SUMMIT. (b) Bronchiectasis-SUMMIT. (c) Cancer-SUMMIT. (d) Multi-Disease-RadChest.** First 100 epochs of training and validation loss curves are shown for TANGERINE and comparison models across example fine-tuning tasks. TANGERINE consistently attains a lower validation loss, indicating increased robustness and generalisation capability. While comparison models converge in terms of training loss, their validation loss fails to decrease suggesting the models fail to learn pattens which are generalisable beyond the training set. This pattern highlights TANGERINE's ability to achieve more stable performance beyond the training set, maintaining better generalisation compared to comparison models. The dotted line indicates TANGERINE optimal epoch for early stopping based on lowest attained validation loss.

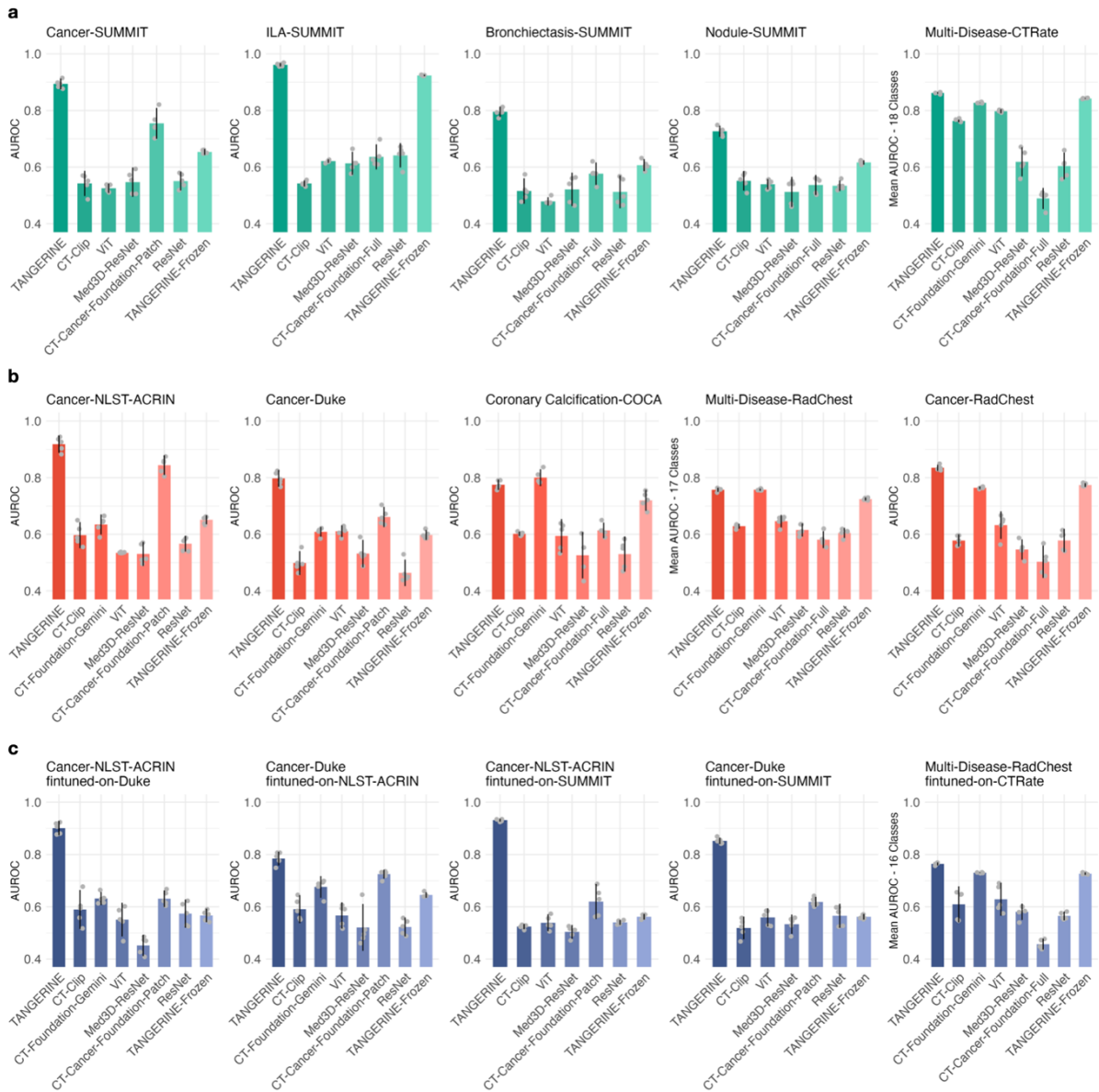

**Extended Data. 8 | Performance on lung disease classification with TANGERINE embeddings.** **(a) Pretrain-seen-distribution**: Models fine-tuned and tested on datasets seen during pretraining TANGERINE. **(b) Pretrain-unseen-distribution**: Models fine-tuned and tested on datasets not seen during pretraining. **(c) Domain generalisation**: Models fine-tuned on one dataset and evaluated on a distinct target dataset unseen during pretraining or finetuning. Each model was trained with five random seeds; error bars show 95% confidence intervals, and bar centres indicate mean AUROC. Use of frozen TANGERINE embeddings (TANGERINE-Frozen) yields strong performance relative to comparison models, with further improvements achieved through end-to-end fine-tuning (TANGERINE).

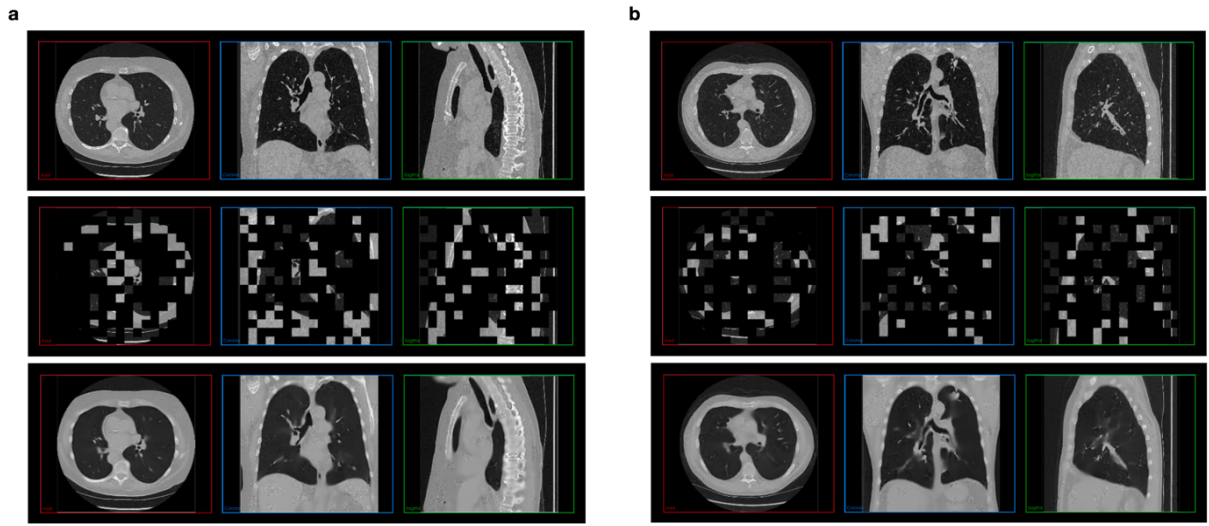

**Extended Data Fig. 9 | Example reconstructions from the masked autoencoding pretraining task.** Representative axial, coronal, and sagittal CT slices before masking (top row), after random masking (middle row), and after reconstruction by the TANGERINE encoder-decoder model (bottom row). Despite heavy masking across multiple anatomical planes, the model accurately reconstructs key lung structures, including bronchi, blood vessels, and pleural boundaries, highlighting the model's capacity to learn spatially coherent representations critical for downstream thoracic disease tasks.

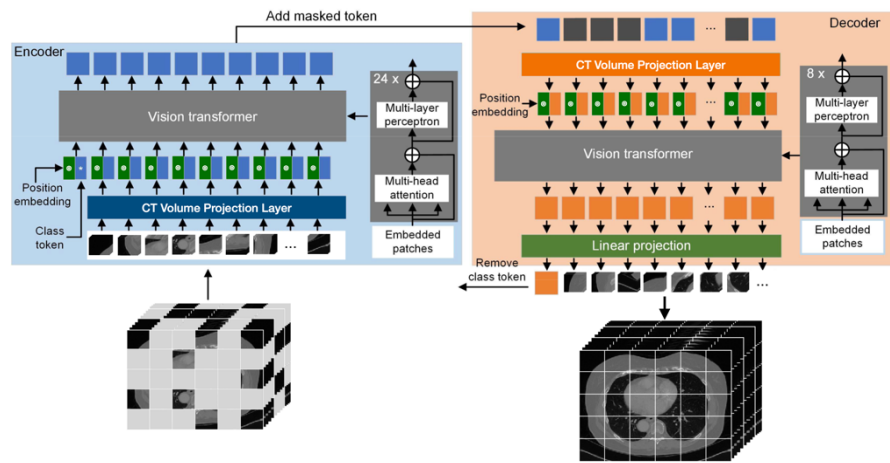

**Extended Data Fig. 10 | Overview of the masked autoencoder (MAE) framework used for pretraining TANGERINE.** Input CT volumes are divided into 3D patches, with a random subset masked prior to encoding. The encoder processes only the visible (unmasked) patches, reducing memory and compute requirements. Masked tokens are reintroduced before the decoder, which reconstructs the original volume from both masked and unmasked representations. Positional embeddings and CT-specific projection layers are used to preserve spatial context throughout the architecture.

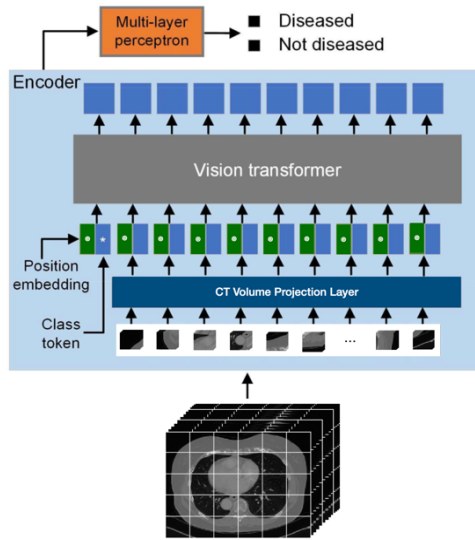

**Extended Data Fig. 11 | Downstream fine-tuning architecture for disease classification.**
During fine-tuning, only the pretrained encoder is retained from the MAE framework. A multi-layer perceptron (MLP) is added on top of the encoder's output to perform binary classification (e.g., diseased vs. not diseased). Input CT volumes are split into 3D patches, embedded via a CT-specific projection layer, and passed through a vision transformer with positional embeddings and a class token. The model is then fine-tuned end-to-end using labelled data for disease-specific tasks.